\documentstyle[11pt,epsf,cite,A4]{article}

\input epsf
\newcommand{\nix}[1]{}
\begin{document}

\begin{center}
{\Large {\bf{Spin Photocurrents in Quantum Wells\\
review part I, (part II: cond-mat/one of the next numbers) }}}
\normalsize

\vspace{5mm}
{\large {\bf{Sergey D. Ganichev and Wilhelm Prettl }}}\\
\vspace{5mm}
Fakult\"{a}t f\"{u}r Physik,
Universit\"at Regensburg, 93040 Regensburg, Germany,\\
\vspace{5mm}
\subsection*{ABSTRACT}
\end{center}
\vspace{0.2mm}
%
\hspace{0.3in}
%
Spin photocurrents generated by homogeneous optical excitation
with circularly polarized radiation in quantum wells (QWs)  are
reviewed. The absorption of circularly polarized light results in
optical spin orientation due to the transfer of the angular
momentum of photons to electrons of a two-dimensional electron gas
(2DEG). It is shown that  in quantum wells belonging to one of the
gyrotropic crystal classes a non-equilibrium spin polarization of
uniformly distributed electrons causes a directed motion of
electron in the plane of the QW. A characteristic feature of this
electric current, which occurs in unbiased samples, is that it
reverses its direction upon changing the radiation helicity from
left-handed to right-handed and vice versa.

Two microscopic mechanisms are responsible for the occurrence of
an electric current  linked to a uniform spin polarization in a
QW: the spin polarization induced circular photogalvanic effect
and the spin-galvanic effect. In both effects the current flow is
driven by an asymmetric distribution of spin polarized carriers in
{\boldmath$k$}-space of systems with lifted spin degeneracy due to
{\boldmath$k$}-linear terms in the Hamiltonian. Spin photocurrents
provide  methods  to investigate  spin relaxation and to conclude
on the in-plane symmetry of QWs. The effect can also be utilized
to develop fast detectors to determine the degree of circular
polarization of a radiation beam.  Furthermore spin photocurrents
at infrared excitation were used to demonstrate and investigate
monopolar spin orientation of free carriers.
\vspace{15mm}
%

\tableofcontents
\newpage
\begin{center}
{\bf {\Large please replace this page by the table of contents of
part II}}
\end{center}
\newpage
\vspace{10mm}

\section{Introduction}

The spin of electrons and holes in solid state systems is an
intensively studied quantum mechanical property showing a large
variety of interesting physical phenomena. Lately, there is much
interest in the use of the spin of carriers in semiconductor
heterostructures together with their charge to realize novel
concepts like spintronics and several schemes of quantum
computation (for review see~\cite{spintronicbook02}).  The
necessary conditions to realize spintronic devices are high spin
polarizations in low dimensional structures and large
spin-splitting of subbands in {\boldmath $k$}-space due to
{\boldmath $k$}-linear terms in the Hamiltonian. The latter is
important for the ability to manipulate spins with an external
electric field by the Rashba effect~\cite{Bychkov84p78}.
Significant progress has been achieved recently in the injection
of spin polarized electrons (or holes), in demonstrating the
Rashba splitting and also in using the splitting for manipulating
the spins~\cite{spintronicbook02}.

One of the most frequently used and  powerful methods of
generation and investigation of spin polarization is optical
orientation~\cite{Meier}. Optical generation of an unbalanced spin
distribution in a semiconductor may lead to a spin photoconductive
effect as well as to a spin photocurrent. In the spin
photoconductive effect the optical spin orientation yields a
change of conductivity which, at application of an external
voltage bias, results in a spin polarized
current~\cite{Haegele98p1580,Kikkawa99p139,Hirohata99p5804}. On
the other hand, spin photocurrents reviewed here occur without an
external bias. They are electric currents which are driven by
optically generated spin polarization.

A spin photocurrent was proposed for   the first time
in~\cite{Averkiev83p393} (see also~\cite{Dyakonov71p144}) and
thereafter observed in bulk AlGaAs~\cite{Bakun84p1293}. In these
works  it was shown that an inhomogeneity of the spin polarization
of electrons results in a surface current due to spin-orbit
interaction. A gradient of spin density was obtained by making use
of the strong fundamental absorption of circularly polarized light
at the band edge of the semiconductor. This and other spin
photocurrents caused by inhomogeneous spin orientation will be
briefly outlined in  section~6.

In the present paper we review a new property of the electron spin
in a homogeneous spin-polarized two-dimensional electron gas: its
ability to drive an electric current if some general symmetry
requirements are met. Recently it was demonstrated that an optical
excitation of quantum well structures with circularly polarized
radiation leads to a current whose direction and magnitude depends
on the degree of circular polarization of the incident
light~\cite{APL00}. This effect belongs to the class of
photogalvanic effects which were intensively studied in
semiconductors (for review
see~\cite{Ivchenko80p427,Belinicher80p199,sturman,book}) and
represents a circular photogalvanic effect (CPGE). The CPGE can be
considered as a transfer  of the photon angular momentum into a
directed motion of a free charge carrier. It is an electronic
analog of mechanical systems which transmit rotatory motion into
linear one like a screw thread or a propeller. The circular
photogalvanic effect was independently predicted  by Ivchenko and
Pikus~\cite{Ivchenko78p74} and Belinicher~\cite{Belinicher78p213}
and then observed in bulk tellurium~\cite{Asnin78p74,Asnin79p565}
(for reviews see \cite{sturman,book}). In tellurium the current
arises due to spin splitting
 of the valence band edge at the boundary of the first Brillouin-zone
 (`camel back' structure)~\cite{Averkiev84p397a402}.
While neither bulk zinc-blende structure materials like GaAs and
related compounds nor bulk diamond structure crystals like Si and
Ge allow this effect, in QW structures CPGE is possible due to a
reduction of symmetry.

It was shown in~\cite{PRL01} that in  zinc-blende structure based
QW structures CPGE is caused by optical spin orientation of
carriers in systems with band splitting in {\boldmath$k$}-space
due to {\boldmath$k$}-linear terms in the Hamiltonian. Here
{\boldmath$k$} is the two-dimensional electron wavevector in the
plane of QW. In this case homogeneous irradiation of QWs with
circularly polarized light results in a non-uniform distribution
of photoexcited carriers in {\boldmath$k$}-space due to optical
selection rules and energy and momentum conservation which leads
to a
current~\cite{PRL01,PhysicaE02,ICPS26invited,PRB02,PRB03inv,Golub03p1,Golub03p2}.
The carrier distribution in the real space remains uniform.

Furthermore, a thermalized but spin-polarized electron gas can
drive an electrical current~\cite{Ivchenko89p175,Ivchenko90p550}. Recently
it was demonstrated that a homogeneous spin polarization obtained
by any means, not necessarily optical, yields a current, if the
same symmetry requirements, which allow {\boldmath$k$}-linear
terms in the Hamiltonian, are met~\cite{Nature02}. This phenomenon
is referred to as spin-galvanic effect. While electrical currents
are usually generated by electric or magnetic fields or gradients,
in this case  a uniform non-equilibrium population of electron
spins gives rise to an electric current. The microscopic origin of
the spin-galvanic effect is an inherent asymmetry of  spin-flip
scattering of electrons in systems with removed
{\boldmath$k$}-space spin degeneracy of the band structure. This
effect has been demonstrated by optical spin
orientation~\cite{Nature02,PASPS02sge,PRB03sge} and
therefore also represents a spin photocurrent.

Both spin orientation induced CPGE and the spin-galvanic effect in
QWs occur at one-photon excitation yielding an electric charge
current linked to a spin polarization. However, a pure spin
current may be obtained at simultaneous one- and two-photon
coherent excitation of proper polarization as recently
demonstrated in bulk GaAs~\cite{Bhat00p5432,Stevens02p4382}. This
phenomenon may be attributed  to a  photogalvanic effect where the
reduced symmetry is caused by the coherent two-frequency
excitation~\cite{Entin89p664} which may also occur in QWs~\cite{Magarill01p652}.

Spin photocurrents at homogeneous excitation have been observed in
$n$- and $p$-type quantum wells based on various semiconductor
materials at very different types of optical excitation by
application of several lasers at wavelengths ranging from the
visible to the far-infrared. There is a particular interest in
spin photocurrents generated by infrared radiation because, in
contrast to conventional methods of optical spin orientation using
inter-band transitions~\cite{Meier}, only one type of charge
carriers is excited yielding monopolar spin
orientation
~\cite{Nature02,PhysicaB99multiphoton,PhysicaE01,PRL02,MRS01monopolar,ICPS26,PASPS02monop}.
Therefore infrared  spin orientation allows to study spin
relaxation without electron-hole interaction and exciton formation
at conditions close to the case of electrical spin
injection~\cite{Nature02,PRL02,MRS01monopolar}.
Finally, spin photocurrents have found technical application as
room temperature detectors which allow to determine and monitor the state of
polarization of terahertz radiation with picosecond time resolution~\cite{MRS02detector}.

This paper is organized in the following way: in section~\ref{II}
an overview of mechanisms yielding photocurrents at homogeneous
spin orientation in QWs is given. First the removal of spin
degeneracy  due to spin-orbit interaction in QWs is discussed and
then it is shown that spin splitting in {\boldmath$k$}-space is
the basic reason for different mechanisms of spin photocurrents in
QWs. Section~\ref{III} gives a short account of the experimental
technique. In section~\ref{IV} the experimental results are
presented and discussed in view of the theoretical background.
Section~5 sketches several kinds of spin independent photocurrents
in comparison to spin photocurrents and experimental methods are
introduced allowing to distinguish between spin dependent and spin
independent currents. Finally, in section~6 we present several
mechanisms of spin photocurrents due to inhomogeneities.

\section{Homogeneous spin orientation induced photocurrents}
\label{II}

Light propagating through a semiconductor and acting upon mobile
carriers can generate a {\it dc}\/ electric current without
external bias. The irradiated sample represents a current source.
Here we consider photocurrents which appear due to optically
induced homogeneous spin orientation of carriers in homogeneous
samples. The microscopic origin of these currents is the
conversion of spin polarization of carriers into directed motion.
The fingerprint of spin photocurrents is their dependence on the
helicity of the radiation field. The current reverses its
direction by switching the polarization of light from right-handed
circular to left-handed circular and vice versa. The experimental
data can be described by simple analytical expressions derived
from a phenomenological theory which shows that the effect can
only be present in gyrotropic media. This requirement rules  out
effects depending on the helicity of the radiation field  in bulk
non-optically active materials like bulk zinc-blende structure and
diamond structure crystals.  The reduction of dimensionality as
realized in QWs makes  spin photocurrents possible. The effect is
quite general and has so far been observed in
GaAs~\cite{APL00,PRL01,Nature02,PRL02},
InAs~\cite{PRL01,Nature02,PASPS02sge}, BeZnMnSe~\cite{DPG01} QW
structures, and  in asymmetric SiGe QWs~\cite{PRB02}.

On a microscopical level spin photocurrents are the result of spin
orientation in systems with {\boldmath$k$}-linear terms in the
Hamiltonian which also may occur in gyrotropic media only. In
general, two mechanisms contribute to spin photocurrents:
photoexcitation and scattering of photoexcited carriers.  The
first  is spin orientation induced circular photogalvanic effect
which is caused by the asymmetry of the momentum distribution of
carriers excited in optical transitions which are sensitive to the
light circular polarization due to selection rules~\cite{PRL01}.
The second mechanism is the spin-galvanic effect which is a result
of spin relaxation. In general this effect does not need optical
excitation but may also occur due to optical spin
orientation~\cite{Nature02}. The current caused by CPGE is spin
polarized and decays with the momentum relaxation time of free
carriers whereas the spin-galvanic effect induced current is
unpolarized but decays with the spin relaxation time.

\subsection{Removal of spin degeneracy}
\label{IIA}

\subsubsection{{\boldmath$k$}-linear terms in the effective Hamiltonian}
\label{IIA1}

Quantum phenomena in semiconductors  are highly sensitive to
subtle details of the carrier energy spectrum so that even a small
spin splitting of energy  bands  may result in measurable effects.
Spin dependent terms linear in the wavevector {\boldmath$k$} in
the effective Hamiltonian remove the spin degeneracy in
{\boldmath$k$}-space of the carrier spectrum. The presence of
these terms  in QWs gives rise to spin photocurrents, yields
beating patterns  in Shubnikov–-de--Haas
oscillations~\cite{Bychkov84p78,Stein83p130,Stormer83p126},
determines spin relaxation in
QWs~\cite{spintronicbook02,Dyakonov86p110,Averkiev02pR271},
results in spin-polarized
tunnelling~\cite{Voskoboynikov98p15397,Silva99p15583,Koga02p126601,PRB03tun},
and allows the control of spin orientation by external
fields~\cite{spintronicbook02,Bychkov84p78,Kalevich90p230,Nitta97p1335,Lu98p1282,Heida98,Hu99p7736,Salis01p619,Smet02p281}.

In the general case, the terms linear in {\boldmath$k$} appear
because the symmetry of heterostructures is lower than the
symmetry of the corresponding bulk materials. Spin degeneracy of
electron bands in semiconductors and subbands of
 heterostructures results because of simultaneous presence of time reversal and
 spatial inversion symmetry. In the present case of low dimensional heterostructures
 and quantum wells, the spatial inversion symmetry is broken. However, in order to obtain
 spin photocurrents depending on the helicity of  radiation and  spin orientation, inversion
 asymmetry is a necessary, but not a sufficient condition. As a matter of fact, the materials must
 belong to one of the gyrotropic crystal classes
which have second rank pseudo-tensors
 as invariants. As a consequence spin dependent
  {\boldmath$k$}-linear terms caused by spin-orbit interaction appear in
 the electron Hamiltonian leading to a splitting of electronic subbands in {\boldmath$k$}-space.
 As long as the time reversal  symmetry is not broken by the application of an external magnetic field,
 the degeneracy of Kramers doublets is not lifted so that still
 $\varepsilon(\mbox{\boldmath$k$},\uparrow ) = \varepsilon(-\mbox{\boldmath$k$},\downarrow )$.
 Here $\varepsilon$ is the electron energy and the arrows indicate the spin orientation.

The principal sources of {\boldmath$k$}-linear terms in the band
structure of QWs are the bulk inversion asymmetry (BIA) of
zinc-blende structure crystals and possibly a structural inversion
asymmetry (SIA) of the low dimensional quantizing structure
(see~\cite{spintronicbook02,Bychkov84p78,PRB02,Dyakonov86p110,Averkiev02pR271,Voskoboynikov98p15397,Silva99p15583,Koga02p126601,PRB03tun,Kalevich90p230,Nitta97p1335,Lu98p1282,Heida98,Hu99p7736,Salis01p619,Smet02p281,Vasko79p994,Roessler84,Cardona88p1806,Lommer88,Luo88p10142,Das89,Luo90p7685,Ivanov91p493,Dresselhaus92,Silva92,Santos94p432,Andrada94p8523,Jusserand95p4707,Pfeffer95pr14332,Ivchenko96p5852,Engels97pr1958,Andrada97p16293,Grundler00p6074,Hammar00,Majewski01,Wilamowski02p195315,Tarasenko02p552}
and references therein). In addition an interface inversion
asymmetry (IIA) may yield {\boldmath$k$}-linear terms caused by
non-inversion symmetric bonding of atoms at heterostructure
interfaces~\cite{spintronicbook02,PRB02,Krebs96p1829,Guettler98pR10179,Krebs98p5770,Toropov00p035302,Olesberg01p201301,Roessler02p313,Golub03bialike}.

BIA induces {\boldmath$k$}-linear terms in the 2D Hamiltonian,
known as Dresselhaus terms, due to the absence of an inversion
center in the bulk crystal. The Dresselhaus terms originate from
the {\boldmath$k$}-cubic terms in the Hamiltonian of a bulk
material~\cite{Dresselhaus55p580}. Averaging these cubic terms
along the quantization axis in the case of low subband filling
with carriers gives rise to the terms linear in {\boldmath$k$}.
These terms are present in QWs based on zinc-blende structure
material and are absent in SiGe heterostructures. IIA may
occur in zinc-blende structure based QWs~\cite{Krebs96p1829,Guettler98pR10179,Krebs98p5770,Toropov00p035302,Olesberg01p201301,Roessler02p313}
were the well and the cladding have different compositions of both
anions and cations like an InAs/GaSb QWs as well as in
SiGe~\cite{PRB02,Golub03bialike}. IIA yields BIA-like terms in the
effective Hamiltonian~\cite{PRB02,Golub03bialike}, thus on a
phenomenological level a separation between BIA and IIA is not
necessary.

The SIA contribution to the removal of spin degeneracy is caused
by the intrinsic he\-terostructure asymmetry which needs not  to be
related to the crystal lattice. These \mbox{{\boldmath$k$}-linear} terms
in the Hamiltonian were first recognized by Rashba and are called
Rashba terms~\cite{Bychkov84p78,Rashba60p1109}. SIA may arise from
different kinds of asymmetries of heterostructures like
non-equivalent normal and inverted interfaces, asymmetric doping
of QWs, asymmetric shaped QWs, external or built-in electric
fields etc. and may also exist in QWs prepared from materials with
inversion symmetry like Si and Ge~\cite{PRB02,Roessler02p313}. It
is the SIA term which allows control of spin polarization by
externally applied electric fields~\cite{Bychkov84p78}. Therefore
these spin-orbit coupling terms are important for spintronics and
in particular for the spin transistor~\cite{Datta90p665}.

In the unperturbed symmetric case we will assume a doubly degenerated subband.
Then the spin-orbit coupling in the non-symmetric structure   has the  form
\begin{eqnarray}
\hat{H}^\prime = \sum_{lm} \beta_{lm}\sigma_lk_m \label{equ1}
\end{eqnarray}
where $\beta_{lm}$ is a second rank
pseudo-tensor and $\sigma_l$ are the Pauli-matrices.
The Pauli matrices occur here because of time reversal symmetry.

In Eq.~(\ref{equ1}) BIA, IIA and SIA can be distinguished by
decomposing $\sigma_lk_m$ into a symmetric and an anti-symmetric
product:
\begin{eqnarray}
\sigma_lk_m = \{\sigma_l,k_m\} +  [\sigma_l,k_m] \label{equ2}
\end{eqnarray}
with the symmetric term
\begin{eqnarray}
\{\sigma_l,k_m\} =\frac{1}{2}\left(\sigma_lk_m +
\sigma_mk_l\right) \label{equ3}
\end{eqnarray}
and the anti-symmetric term
\begin{eqnarray}
[\sigma_l,k_m] =\frac{1}{2}\left(\sigma_lk_m - \sigma_mk_l\right).
\label{equ4}
\end{eqnarray}
Now the perturbation can be written as:
\begin{eqnarray}
\hat{H}^\prime = \sum_{lm}\left(\beta^s_{lm}\{\sigma_lk_m\} +
\beta^a_{lm} [\sigma_lk_m]\right) \label{equ5}
\end{eqnarray}
where $\beta^s_{lm}$ and $\beta^a_{lm}$ are symmetric and
anti-symmetric pseudo-tensors projected out of the full tensor by
the symmetric and anti-symmetric products of $\sigma_lk_m$,
respectively. The symmetric term describes BIA as well as possible
IIA-terms  whereas the anti-symmetric term is caused by SIA.

\subsubsection{Spin splitting of energy  bands in  zinc-blende structure based QWs}
\label{IIA2}

The pseudo-tensor  $\beta_{lm}$ as a material property must
transform after the identity representation of the  point group
symmetry of the quantum well. The point group is determined by the
crystallographic orientation  and the profile of growth and doping
of QWs. The three point groups $D_{2d}$, $C_{2v}$ and $C_s$ are
particularly relevant for zinc-blende structure based
QW~\cite{book}. Hereafter the Sch\"onflies notation is used to
label the point groups. In the international notation they are
labelled as $\bar{4}2m$, $mm2$ and $m$, respectively. The $D_{2d}$
point-group symmetry corresponds to  perfectly grown
(001)-oriented  QWs with symmetric doping. In such QWs only BIA
and IIA terms may exist. The symmetry of (001)-grown QWs reduces
from $D_{2d}$ to $C_{2v}$ if an additional asymmetry is present
due to e.g. non-equivalent interfaces, asymmetric growth profiles,
asymmetric doping etc. resulting in SIA. The relative strength of
BIA, IIA and SIA depends on the structure of the quantum well. In
structures of strong growth direction asymmetry like
heterojunctions  the SIA term  may be larger than that of BIA and
IIA. The last point group is $C_s$, which contains only two
elements, the identity and one mirror reflection plane. It is
realized for instance in (113)- and miscut (001)- oriented
samples.

The non-zero components of the pseudo-tensor $\beta_{lm}$ depend
on the symmetry and the coordinate system used. For
(001)-crystallographic orientation grown QWs of $D_{2d}$ and
$C_{2v}$ symmetry  the tensor elements are given in the coordinate
system $(xyz)$ with $x\parallel [1\bar{1}0]$, $y\parallel [110]$,
$z\parallel [001]$. The coordinates $x$ and $y$ are in the
reflection planes of both point groups perpendicular to the
principle two fold axis; $z$ is along the growth direction normal
to the plane of the QW. In $D_{2d}$ the pseudo-tensor $\beta_{lm}$
is  symmetric, $\beta_{lm}=\beta^s_{lm}$. In the above coordinate
system there are two non-zero components $\beta_{xy}$ and
$\beta_{yx}$ with $\beta_{yx}=\beta_{xy}= \beta^s_{yx}$. For
zinc-blende structure type crystals it has been shown that the BIA
and IIA terms in the Hamiltonian have the same form, thus, IIA
enhances or reduces the strength of BIA-like term.

Therefore we obtain\footnote{For coordinates along cubic axes,
$x\parallel$[100]and  $y\parallel$[010], we have non-zero  components
$\beta_{xx}$ and $\beta_{yy}$ with $\beta_{yy} =-\beta_{xx}$ which yields
$ \hat{H}^\prime = \hat{H}_{BIA}+\hat{H}_{IIA} = \beta_{xx} (\sigma_xk_x - \sigma_y k_y)$. }%
\begin{eqnarray}
\hat{H}^\prime = \hat{H}_{BIA}+\hat{H}_{IIA} = \beta^s_{xy}
(\sigma_xk_y +\sigma_y k_x) \label{equ6}
\end{eqnarray}

In $C_{2v}$ the tensor $\beta_{lm}$ is non-symmetric yielding additional terms in $\hat{H}^\prime$
caused by SIA so that now $\hat{H}^\prime = \hat{H}_{BIA}+\hat{H}_{IIA} + \hat{H}_{SIA}$. The form
of $\hat{H}_{BIA}$ and $\hat{H}_{IIA}$ remains unchanged by the reduction of symmetry from $D_{2d}$ to $C_{2v}$.
The SIA term in $C_{2v}$ assumes the form:
\begin{eqnarray}
\hat{H}_{SIA} = \beta^a_{xy} (\sigma_xk_y -\sigma_y k_x).
\label{equ7}
\end{eqnarray}
It is clear that the form of  this term is independent of the
orientation of cartesian coordinates in the plane of the QW. The
strength of spin splitting was experimentally derived e.g. from
beatings of Shubnikov-de-Haas oscillations in various
III-V-compound based
QWs~\cite{Bychkov84p78,Stein83p130,Stormer83p126}. It has been
found to be in the range of 10$^{-10}$  and 10$^{-9}$ eV$\cdot$cm
and was attributed to structural inversion asymmetry.

The point group $C_s$ is discussed by example of (113)-orientation
grown QWs because they are available and spin photocurrents in
them have been intensively  investigated so far\footnote{Miscut
(001)-oriented samples investigated in~\cite{APL00} also have
$C_s$ symmetry.}. In this case we use the coordinates $x^\prime =
x\parallel [1\bar{1}0]$, as above, $y^\prime \parallel
[33\bar{2}]$, $z^\prime \parallel [113]$. Direction $x$ is normal
to the reflection plane, the only non-identity symmetry element of
this group, and $z^\prime $ is along the growth direction. The
reduction of the symmetry to $C_s$ results in an additional term
in the Hamiltonian:
\begin{eqnarray}
\hat{H}^\prime = \beta_{z^{\prime} x} \sigma_{z^{\prime}}k_x.
\label{equ8}
\end{eqnarray}

\begin{figure}
\centerline{\epsfxsize 140mm \epsfbox{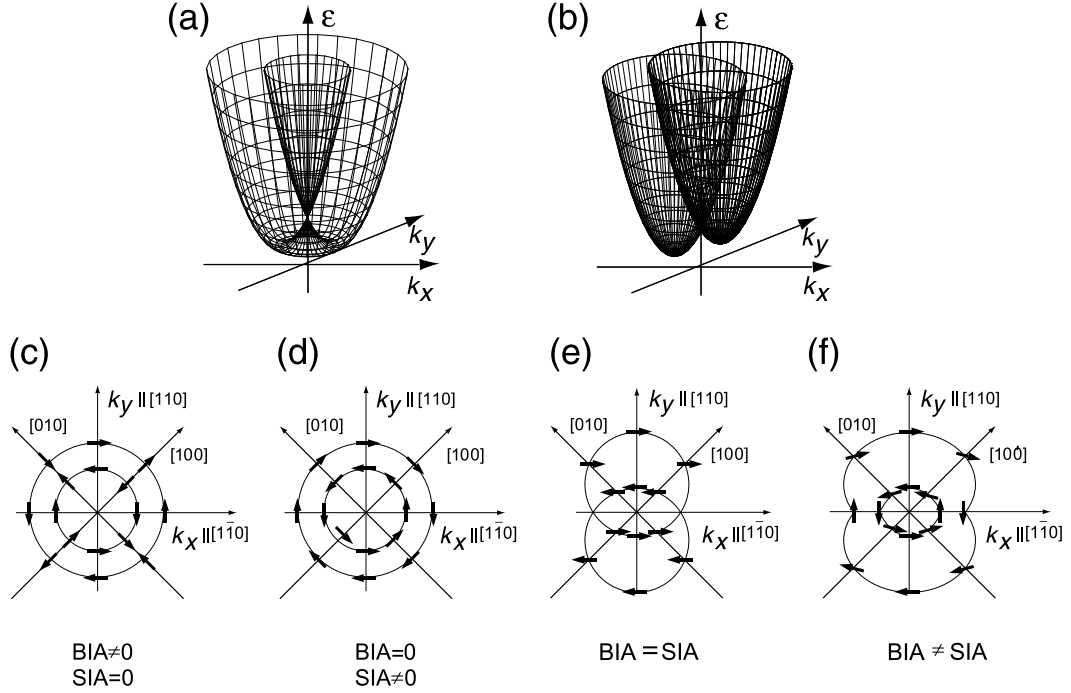  }}
\caption{ Schematic 2D  band structure with {\boldmath$k$}-linear
terms for $C_{2v}$ symmetry. The energy $\varepsilon$ is plotted
as a function of $k_x$ and $k_y$ in (a) with only one type of
inversion asymmetry, BIA or SIA, respectively and in (b) for equal
strength of the BIA and SIA terms in the Hamiltonian.
The bottom plate shows the distribution of spin orientations at
the 2D Fermi energy for different strength of the BIA and SIA
terms. }
\label{fig01}
\end{figure}

In order to illustrate  band structures with a
{\boldmath$k$}-linear term in Fig.~\ref{fig01} we plotted  the
energy $\varepsilon$ as a function of $k_x$ and $k_y$ for $C_{2v}$
symmetry. The upper plate of Fig.~\ref{fig01} shows the band
structure with only one type of inversion asymmetry, BIA or SIA
(Fig.~\ref{fig01}a) and the band structure at equal strength of
the BIA (including IIA) and SIA (Fig.~\ref{fig01}b). In the
illustration we assume positive coefficients $\beta_{lm}^a,
\beta_{lm}^s \geq 0$. In the case of BIA only ($\beta^a_{lm} = 0$)
or SIA only ($\beta^s_{lm}=0$) the band structure is the result of
the revolution around the energy axis of two parabolas
symmetrically displaced with respect to {\boldmath$k$} = 0. A
constant energy surface is a pair of concentric circles, however,
the spins are  oriented differently for BIA and SIA.
 The
distribution of spin orientation in {\boldmath$k$}-space, obtained
by the procedure  of~\cite{Silva92}, is indicated by arrows in the
bottom plate of Fig.~\ref{fig01}. The distribution of spins for a
pure BIA term is shown in Fig.~\ref{fig01}c. If only the SIA term
is present (Fig.~\ref{fig01}d) then the spins  are always oriented
normal to the wavevector {\boldmath$k$}. This is a consequence of
the vector product in the Rashba spin-orbit
interaction~\cite{Bychkov84p78}. If the strengths of BIA and SIA
are the same then the 2D band structure consists of two revolution
paraboloids with revolution axes symmetrically shifted  in
opposite direction with respect to {\boldmath$k$} = 0
(Fig.~\ref{fig01}b). Now the spins are oriented along $\pm k_x$ as
shown in Fig.~\ref{fig01}e. In Fig.~\ref{fig01}f we have shown a
constant energy surface and direction of spins for $\beta_{lm}^a
\neq \beta_{lm}^s$.

Finally we briefly discuss QWs prepared on SiGe. As both Si and Ge
possess inversion centers there is no  BIA, however both IIA, with
BIA-like form of the Hamiltonian, and SIA may lead to
{\boldmath$k$}-linear
terms~\cite{PRB02,Wilamowski02p195315,Roessler02p313,Golub03bialike}.
The symmetry of Si/(Si$_{1-x}$Ge$_x$)$_n$/Si QW depends on the
number~$n$ of the monoatomic layers  in the well. In the case of
(001)-crystallographic orientation grown  QW structures with an
even number~$n$, the symmetry of  QWs is $D_{2h}$ which is
inversion symmetric and does not yield {\boldmath$k$}-linear
terms. An odd number of~$n$, however, interchanges the
$[1\bar{1}0]$ and $[110]$ axes of the adjacent barriers and
reduces the symmetry to $D_{2d}$ with the same implication treated
above for zinc-blende structure QWs~\cite{PRB02}.

\subsection{Circular photogalvanic effect}
\label{IIB}

\subsubsection{Microscopic model}
\label{IIB1}

\begin{figure}
\centerline{\epsfxsize 120mm \epsfbox{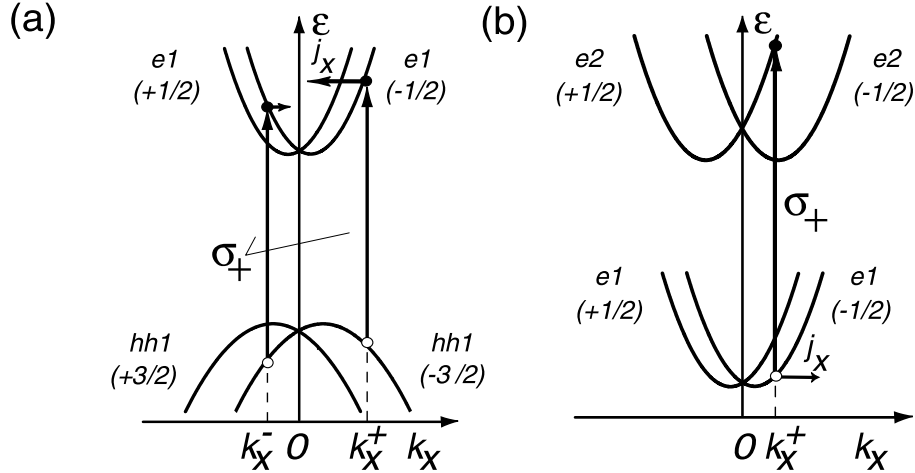  }}
\caption{Microscopic picture of  spin orientation induced CPGE at
direct transitions in C$_s$ point group taking into account the
splitting of subbands in {\boldmath$k$}-space. $\sigma_+$
excitation induces direct transitions (solid arrows) (a) between
valence and conduction band (from $hh${\it 1} ($m_s = -3/2$) to
$e${\it 1} ($m_s=-1/2$) and (b) between size quantized subbands in
the conduction band (from $e${\it 1} ($m_s = -1/2$) to  $e${\it 2}
($m_s=+1/2$)). Spin splitting together with optical selection
rules results in an unbalanced occupation of the positive $k_x^+$
and negative $k_x^-$ states yielding a spin polarized
photocurrent. For $\sigma_-$ excitation both the spin orientation
of the charge carriers and the current direction get reversed. In
each plate arrows indicate the current due to an unbalance of
carriers. Currents are  shown for  one subband only. }
\label{fig02}
\end{figure}

{\it Inter-band transitions:} The spin orientation induced
circular photogalvanic effect  is most easily conceivable for both
$n$- and $p$-type materials from the schematic band structure
shown in Fig.~\ref{fig02}a~\cite{PRL01}. We assume direct
inter-band transitions in a QW of $C_s$ symmetry. For the sake of
simplicity we take into account a one dimensional band structure
consisting  only of the lowest conduction subband $e${\it 1} and
the highest heavy-hole subband $hh${\it 1}.   The splitting in the
conduction band is  given by $\varepsilon_{e{\it 1}, \pm
1/2}(\mbox{\boldmath$k$}) = [(\hbar^2 k^2_x/2 m_{e{\it 1}}) \pm
\beta_{e{\it 1}} k_{x } + \varepsilon_g]$ and in the valence band
by $\varepsilon_{hh{\it 1}, \pm 3/2}(\mbox{\boldmath$k$}) =
-[(\hbar^2 k^2_x/2 m_{hh{\it 1}}) \pm \beta_{hh{\it 1}} k_{x }$],
where $\varepsilon_g$ is the energy gap.

%

For absorption of circularly polarized radiation of photon energy
$\hbar\omega$ energy and momentum conservation allow transitions
only  for  two  values of $k_x$. Due to selection rules the
optical transitions occur from $m_s=-3/2$ to $m_s=-1/2$ for right
handed circular polarization ($\sigma_{+}$) and from $m_s = 3/2$
to $m_s = 1/2$ for left handed circular polarization
($\sigma_{-}$). Here $m_s$ are the spin quantum numbers of the
electron states. The corresponding transitions for e.g.
$\sigma_{+}$ photons  occur at
\begin{equation}
k_{x }^{\pm} = + \frac{\mu}{\hbar^2} (\beta_{e{\it
1}}+\beta_{hh{\it 1}} )\pm \sqrt{ \frac{\mu^2}{\hbar^4}
(\beta_{e{\it 1}} +\beta_{hh{\it 1}} )^2 + \frac{2
\mu}{\hbar^2}\left( \hbar \omega - \varepsilon_g \right)},
\label{equ9}
\end{equation}
 and are shown in
Fig.~\ref{fig02}a by the solid  vertical arrows. Here $\mu =
(m_{e{\it 1}}\cdot m_{hh{\it 1}})/(m_{e{\it 1}}+m_{hh{\it 1}})$ is
a reduced mass. The `center of mass' of these transitions is
shifted from the point \mbox{$k_{x}=0$} by  $(\beta_{e{\it
1}}+\beta_{hh{\it 1}}) (\mu / \hbar^2)$. Thus the sum of the
electron velocities in the excited states in the conduction band,
$v_{e{\it 1}}= \hbar (k_{x }^{-} + k_{x }^{+}-2k_x^{min}) /
m_{e{\it 1}} = 2/[\hbar( m_{e{\it 1}} + m_{hh{\it
1}})]\cdot(\beta_{hh{\it 1}}m_{hh{\it 1}}-\beta_{e{\it 1}}m_{e{\it
1}}) $, is non-zero. The contributions of $k_{x}^{\pm}$
photoelectrons to the current do not cancel each other except in the
case of $\beta_{e{\it 1}}m_{e{\it 1}}=\beta_{hh{\it 1}}m_{hh{\it
1}}$ which corresponds to an equal splitting of the conduction and the valence band.
We note
that the group velocity is obtained taken into account that
$k_{x}^{\pm}$ are to be counted from the conduction subband minima
$k_x^{min}$ because the current is caused by the difference  of the
group velocities within the subband. The same consideration
applies for holes in the initial states in $hh${\it 1}.
Consequently, a spin polarized net current in the $x$ direction
results. Changing the circular polarization of the radiation  from
$\sigma_{+}$ to $\sigma_{-}$ reverses  the current because the
`center of mass' of these transitions is now shifted to
$-(\beta_{e{\it 1}}+\beta_{hh{\it 1}}) (\mu / \hbar^2)$.

{\it Inter-subband transitions:} In the longer wavelength range,
infrared or far-infrared,  the current is caused by inter- or
intra-subband transition. For direct transition between size
quantized  states in the valence or conduction band, for example
like  $e${\it 1} and $e${\it 2} in $n$-type materials, the model
is very similar to inter-band transitions discussed
above~\cite{PRB03inv}. In Fig.~\ref{fig02}b we shortly sketch this
situation for QWs of $C_{s}$ symmetry. The $\sigma_{z^\prime} k_x$
contribution to the Hamiltonian splits the electron spectrum into
spin sub-levels with the spin components $m_s=\pm 1/2$ along the
growth direction $z^\prime$.   As a result of optical selection
rules right-handed circular polarization under normal incidence
induces direct optical transitions between the subband $e${\it 1}
with spin $m_s=-1/2$ and $e${\it 2} with spin $m_s=+1/2$. For
monochromatic radiation optical transitions  occur only at a fixed
$k_x^+$ where the energy of the incident light matches the
transition energy as is indicated by the arrow in
Fig.~\ref{fig02}b. Therefore optical transitions induce an
imbalance of momentum distribution in both subbands yielding an
electric current in the $x$ direction with contributions from
$e${\it 1} and $e${\it 2}. As in $n$-type QWs the energy
separation between $e${\it 1} and $e${\it 2} is typically larger
than the energy of longitudinal optical phonons
$\hbar\omega_{LO}$, the non-equilibrium distribution of electrons
in $e${\it 2}  relaxes rapidly due to emission of phonons. As a
result, the contribution of the $e${\it 2} subband to the electric
current vanishes. Thus the magnitude and the direction of the
current is determined by the group velocity and the momentum
relaxation time $\tau_p$ of photogenerated  holes in the initial
state of the resonant optical transition in the $e${\it 1} subband
with $m_s=-1/2$.

%
\begin{figure}
\centerline{\epsfxsize 50mm \epsfbox{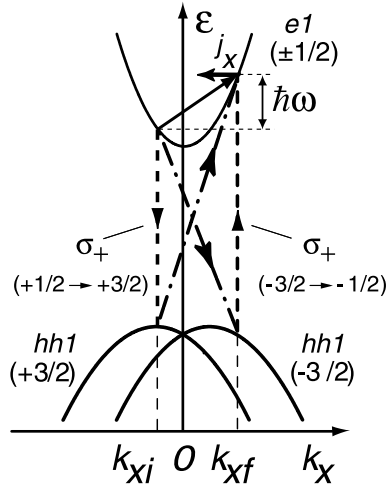  }}
\caption{Microscopic picture describing the origin of spin
orientation induced CPGE at indirect (Drude) transitions in C$_s$
point group samples. }
\label{fig03}
\end{figure}

{\it Intra-subband transitions (Drude absorption):} Now we
consider indirect intra-subband transitions. This situation is
usually realized in  the far-infrared range where the photon
energy is not high enough to excite direct inter-subband
transitions. Due to energy and momentum conservation intra-subband
transitions can only occur by absorption of a photon and
simultaneous absorption or emission of a phonon. This process is
described by virtual transitions involving intermediate states and
will be discussed in more detail in section~4.3.2. It can be shown
that transitions  via intermediate states  within one and the same
subband  do not yield spin orientation and do not contribute to
the spin photocurrent. However, spin selective indirect optical
transitions excited by  circularly polarized light with both
initial and final states in the conduction band can  generate a
spin current  if virtual processes involve intermediate states in
different subbands~\cite{PRL01}. Fig.~\ref{fig03} sketches the
underlying mechanism for $\sigma_+$ polarization. For the sake of
simplicity only the spin splitting of the valence band is taken
into account. The two virtual transitions shown represent
excitations which, for $\sigma_+$ helicity, transfer electrons
from states with negative $k_x$ to states with positive $k_x$. The
current resulting from a free electron transition (solid arrow) in
the conduction band $e${\it 1} occurs due to transitions involving
intermediate states in the valence subbands. Two representative
virtual transitions for $\sigma_+$ excitation are illustrated in
Fig.~\ref{fig03}. One is an optical transition from $m_s = +1/2$
to $m_s = +3/2$ (dashed line, downward arrow) and a transition
involving a phonon from $m_s = +3/2$ back to the conduction band
(dash-dotted line, upward arrow). The other is a phonon transition
from the conduction band to the $m_s = -3/2$ intermediate state in
$hh${\it 1} and an optical transition from $m_s = -3/2$ to $m_s =
-1/2$. While the first route depopulates preferentially initial
states of spin $m_s=+1/2$ for $k_{xi} < 0$, the second one
populates preferentially final state of $m_s = -1/2$ states for
$k_{xf} > 0$~\cite{PRL01}. This together with the unbalanced
occupation of the {\boldmath$k$}-space causes  a spin-polarized
photocurrent. Switching the helicity from $\sigma_+$ to $\sigma_-$
reverses the process and results in a spin photocurrent in the
opposite direction.

\subsubsection{Phenomenology}
\label{IIB2}

On the macroscopic  level the CPGE
can be described by the following  phenomenological expression~\cite{book}:
\begin{equation}
j_{\lambda} = \sum_{\mu}\gamma_{\lambda \mu}\:i (\mbox{\boldmath$
E$} \times {\mbox{\boldmath $E$}}^* )_{\mu} \:, \label{equ10}
\end{equation}
\begin{equation}
 i (\mbox{\boldmath$ E$} \times {\mbox{\boldmath $E$}}^* )_{\mu} = \hat{e}_\mu \:E_0^2 P_{circ}
 \label{equ11}
\end{equation}
  where {\boldmath$j$} is the  photocurrent density,
{\boldmath$\gamma$} is a second rank pseudo-tensor,
{\boldmath$E$} is the complex amplitude of the electric field of
the electromagnetic wave,
$E_0$, $P_{circ}$, {\boldmath$\hat{e}$} = {\boldmath$q$}$/q$ and
{\boldmath$q$} are the electric field amplitude, the degree of
circular polarization, the unit vector pointing in the direction
of light propagation and  the light wavevector inside the medium,
respectively. The photocurrent  is proportional to
the radiation helicity $P_{circ}$ and can be observed only under circularly
polarized excitation.
The helicity of the incident radiation
is given by
\begin{equation}
P_{circ}= \frac{I_{\sigma_+}-I_{\sigma_-}}{I_{\sigma_+} + I_{\sigma_-} } = \sin 2\varphi
\label{equ12}
\end{equation}
where $\varphi$ is the phase angle between the $x$ and $y$
component of the electric field vector. $P_{circ}$  varies from
$-1$ (left-handed circular, $\sigma_-$) to $+1$ (right-handed
circular, $\sigma_+$).

In general, in addition to the circular photogalvanic current
given in Eq.~(\ref{equ10}), two other photocurrents can be
simultaneously present, namely the linear photogalvanic effect
(LPGE) and the photon drag effect. Both effects were observed in
low dimensional structures. They do not require spin orientation
and will be summarized in   section~5.

As a result of tensor equivalence the second rank pseudo-tensor
{\boldmath$\gamma$} is subjected to the same symmetry restriction
as {\boldmath$\beta$} responsible for the {\boldmath$k$}-linear
terms in the Hamiltonian,  discussed in detail above.  Thus,
$\gamma_{lm}$ depends in the same way like $\beta_{lm}$ on
the symmetry and the coordinate system. In the various crystal classes
being of importance here, the same elements of $\beta_{lm}$ are
non-zero like those of $\gamma_{lm}$.

In the following we analyze Eq.~(\ref{equ10}) for $D_{2d}$,
$C_{2v}$ and $C_s$ in the coordinate systems $(xyz)$ and $(x
y^\prime z^\prime )$. Due to carrier confinement in growth
direction the photocurrent in QWs has non-vanishing components
only in the plane of a QW.

For the point group $D_{2d}$ the non-zero components of
{\boldmath$\gamma$}  are $\gamma_{x y }$ and $\gamma_{y  x }$ with $
\gamma_{x y } = \gamma_{y  x }$. We denote the only independent
element by $\gamma^{(0)} = \gamma_{x y } $; then the current in a
QW is given by:
\begin{equation}
j_x = \gamma^{(0)}\hat{e}_y E^2_0 P_{circ}  \:,\: \: \: \: \: \:
\: \: \: \: \: j_y = \gamma^{(0)}\hat{e}_x E^2_0  P_{circ} \:.
\label{equ13}
\end{equation}
where $E^2_0$ is the square of the electric field amplitude in
vacuum being proportional to the radiation power  $P$.

Eqs.~(\ref{equ13}) shows that in this configuration we get a
transverse effect if the sample is irradiated along a $\langle
110\rangle$ crystallographic orientation, corresponding to
$\hat{e}_x=1$, $\hat{e}_y =0$  or $\hat{e}_x=0$, $\hat{e}_y =1$.
The current {\boldmath$j$} is perpendicular to the direction of
light propagation {\boldmath$\hat{e}$}. If the radiation is shined
in along a cubic axis $\langle 100\rangle$, with $ \hat{e}_x=
\hat{e}_y =1/\sqrt{2}$, then the current is longitudinal flowing
along the same cubic axis because $j_x = j_y$. Putting all
together, we see from Eqs.~(\ref{equ13}) that rotating
{\boldmath$\hat{e}$} in the plane of the QW counter-clockwise
yields a clockwise rotation of {\boldmath$j$}.

\begin{figure}
\centerline{\epsfxsize 60mm \epsfbox{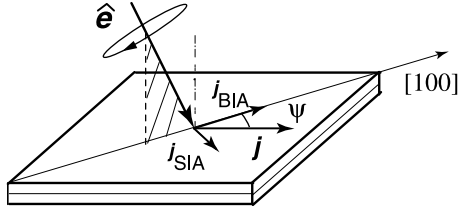  }}
\caption{SIA and BIA- induced circular photogalvanic effect
generated in samples of C$_{2v}$ symmetry under oblique incidence
of circularly polarized light with excitation along [110].}
\label{fig04}
\end{figure}

Reducing the symmetry from $D_{2d}$ to $C_{2v}$, the tensor
{\boldmath$\gamma$} describing the CPGE is characterized by two
independent components $\gamma_{xy}$ and $\gamma_{yx}\neq
\gamma_{xy}$.  We define $\gamma^{(1)} = \gamma_{xy}$ and
$\gamma^{(2)} = \gamma_{yx}$,  then the photocurrent is determined
by
\begin{equation}
j_{x} = \gamma^{(1)} \hat{e}_y E^2_0  P_{circ}  \:,\: \: \: \: \:
\: \: \: \: \: \: j_{y} = \gamma^{(2)} \hat{e}_x E^2_0  P_{circ}.
\label{equ14}
\end{equation}
If {\boldmath$\hat{e}$}  is along  $\langle 110\rangle$ so as
$\hat{e}_x=1$ and $\hat{e}_y=0$ or $\hat{e}_x=0$ and
$\hat{e}_y=1$, then the current again flows  normal to the
light propagation direction. In contrast to $D_{2d}$ symmetry the
strength of the current is different for the radiation propagating
along $x$ or $y$. This is due to the non-equivalence of the
crystallographic axes [1$\bar{1}$0] and [110] because of the
two-fold rotation axis in $C_{2v}$ symmetry. If the sample is
irradiated with {\boldmath$\hat{e}$}
 parallel to $\langle 100 \rangle$ corresponding
to $\hat{e}_x =\hat{e}_y = 1/\sqrt{2}$, the current is neither
parallel nor perpendicular to the light propagation direction (see
Fig.~\ref{fig04}). The current includes an angle $\psi$ with the
$x$-axis given  by $\tan \psi = \gamma_{xy}/\gamma_{yx} =
\beta_{yx}/\beta_{xy}$. The last equation follows from tensor
equivalence.

%

Another conclusion from Eqs.~(\ref{equ13}) and~(\ref{equ14}) is
that in QWs of the higher symmetries $D_{2d}$ and $C_{2v}$ the
photocurrent can only be induced under oblique incidence of
irradiation. For normal incidence {\boldmath$\hat{e}$} is parallel
to [001] and hence the current vanishes as $\hat{e}_x= \hat{e}_y =
0$. In contrast to this result in QWs of  $C_s$ symmetry a
photocurrent also occurs  for normal incidence of the radiation on
the plane of the QW because  the tensor {\boldmath$\gamma$} has an
additional component $\gamma_{x z^{\prime}}$. The current here is
given by
\begin{equation}
j_x = ( \gamma_{x y^{\prime} } \hat{e}_{y^\prime}  + \gamma_{x
z^{\prime} } \hat{e}_{z^\prime}  ) E^2_0  P_{circ}  \:,\: \: \: \:
\: \: \: \: \: \: \: j_{y^\prime} = \gamma_{y^{\prime} x}
\hat{e}_x E^2_0  P_{circ}. \label{equ15}
\end{equation}
At normal incidence, $\hat{e}_x = \hat{e}_{y^\prime} =0 $ and $\hat{e}_{z^\prime} = 1$, the
current  in the QW  flows perpendicular to the mirror
reflection plane of $C_s$ which corresponds to the  $x$ coordinate  parallel to
$[1\bar{1}0]$.

Now we will take a closer look on the dependence of the
photocurrent on the angle of incidence $\Theta_0$ which is
determined by the value of the projection {\boldmath$\hat{e}$} on
the $x$- ($y$-) axis (see Eqs.~(\ref{equ13}) and (\ref{equ14})) or
on the $z^\prime$-axis (Eqs.~(\ref{equ15})). We have for  the
excitation in the  plane of incidence parallel to $(yz)$
 \begin{equation}
\hat{e}_x = t_p t_s \sin \Theta, \label{equ16}
\end{equation}
and in  the plane of incidence parallel to ($y^\prime z^\prime$)
\begin{equation}
\hat{e}_{z^\prime} =  t_p t_s \cos \Theta, \label{equ17}
\end{equation}
where $\Theta$ is the refraction angle defined by $\sin{\Theta} =
\sin{\Theta_0}/\sqrt{\varepsilon^*}$,  $\varepsilon^*$ is the dielectric constant of the QW
material, and  transmission coefficients $t_p$, $t_s$ for linear $p$ and $s$ polarizations after
Fresnel's formula are given by
\begin{equation}
t_p t_s= \frac{4 \cos^2{\Theta_0}} {\Bigl(\cos{\Theta_0} +
\sqrt{\varepsilon^* - \sin^2{\Theta_0}}\Bigr) \Bigl(\varepsilon^*
\cos{\Theta_0} + \sqrt{\varepsilon^* - \sin^2{\Theta_0}}
\Bigr)}\:\:. \label{equ18}
\end{equation}
For $\hat{e}_{y^\prime}$ in the left equation of
Eqs.~(\ref{equ15}) we obtain $\hat{e}_{y^\prime} = t_p t_s \sin
\Theta$ for  the excitation in the  plane of incidence parallel to
$(y^\prime z)$.

The measurement of the CPGE with respect to the angle of incidence
and the crystallographic direction is important to determine the
in-plane symmetry of the QW. Indeed, only in $C_s$ symmetry CPGE
occurs at normal incidence (see Eqs.~(\ref{equ15})), and $D_{2d}$
and $C_{2v}$ symmetries may be distinguished  by excitation along
a $\langle 100 \rangle$ axis, because in this case only $D_{2d}$
does not allow a transverse effect.

\subsubsection{Microscopic theory}
\label{IIB3}

{\it Inter-band transitions:} The microscopic theory of spin
orientation induced CPGE in QWs was worked out for inter-band
excitation in~\cite{PhysicaE02,Golub03p1,Golub03p2} and is briefly
sketched here following ~\cite{PhysicaE02}. We consider the
asymmetry of the momentum distribution of holes excited under
direct inter-band optical transitions in $p$-doped $(113)$-grown
QWs of $C_s$ symmetry. We remind that in this case normal incident
radiation of circular polarization induces a current in  $x$
direction. Let us denote the free hole states in a QW as $\vert
\nu m_s \mbox{\boldmath $k$} \rangle$, where  $\nu$ and $m_s$ are
the hole subband and spin-branch indices, respectively. If only
terms even in {\boldmath$k$} are taken into account in the
effective Hamiltonian of holes, all  hole subbands $(\nu,
\mbox{\boldmath$k$})$ are doubly degenerate. Allowing  terms odd
in {\boldmath$k$} in the Hamiltonian results in a subband spin
splitting so that the hole energy $\varepsilon_{\nu m_s
\mbox{\footnotesize{\boldmath$k$}} }$ becomes dependent on the
spin branch index $m_s$. The photocurrent density is given by a
standard expression
\begin{equation}
j_{x} = e \sum_{\nu m_s \mbox{\footnotesize{\boldmath$k$}} }
u_{x}(\nu m_s \mbox{\boldmath$k$} ) f_{\nu m_s
\mbox{\footnotesize{\boldmath$k$}} }\:,
\label{equ19}
\end{equation}
where $e$ is the elementary charge (for holes $e>0$), $u_{x}(\nu
m_s \mbox{\boldmath$k$})$ is the group velocity $\hbar^{-1}
(\partial \varepsilon_{\nu m_s \mbox{\footnotesize{\boldmath$k$}} } /
\partial k_{x})$ and $f_{\nu m_s
\mbox{\footnotesize{\boldmath$k$}} }$ is the non-equilibrium
steady-state distribution function. Note that the energy
$\varepsilon_{\nu m_s \mbox{\footnotesize{\boldmath$k$}} }$ is
invariant and the velocity $ u_{x}(\nu m_s
\mbox{\footnotesize{\boldmath$k$}})$ changes its sign under the
time-inversion operation $K$ transforming a spinor $\hat{\psi}$
into $K \hat{\psi}$ $\equiv $ $i \sigma_y \hat{\psi}$ ($\sigma_y$
is one of the Pauli matrices)~\cite{book}. Therefore only  the
anti-symmetric part of the distribution function $f^{-}_{\nu m_s
\mbox{\footnotesize{\boldmath$k$}} }$ = $( f_{\nu m_s
\mbox{\footnotesize{\boldmath$k$} }} - f_{\nu \bar{m_s}, -
\mbox{\footnotesize{\boldmath$k$}} } ) / 2$  contributes to $j_x$.
Here $\vert \nu \bar{m_s}, - \mbox{\boldmath$k$} \rangle$ is
obtained from $\vert \nu m_s \mbox{\boldmath$k$} \rangle$ by
application of the operator $K$.

In the momentum relaxation time approximation under direct
optical transitions we have
\begin{eqnarray}
j_{x} = e \sum_{\nu' \nu m'_s m_s \mbox{\footnotesize{\boldmath$k$}} } W_{\nu' m'_s,
\nu m_s}(\mbox{\boldmath$k$} , \mbox{\boldmath$e$}
 ) \left[ u_{x}(\nu' m'_s \mbox{\boldmath$k$}
) \tau_p^{(\nu')} -
 u_{x}(\nu m_s \mbox{\boldmath$k$}
) \tau_p^{(\nu)} \right] \:,
\label{equ20}
\end{eqnarray}
where $ \mbox{\boldmath$e$} $ is the light
polarization unit vector and  $\tau_p^{(\nu)}$ is the hole momentum
relaxation time in the subband $\nu$. The probability rate for the
transition $ | \nu m_s \mbox{\boldmath$k$} \rangle \rightarrow |
\nu' m'_s\mbox{\boldmath$k$} \rangle $ is given by  Fermi's golden
rule
\begin{eqnarray}
W_{\nu' m'_s, \nu m_s}(\mbox{\boldmath$k$} ,  \mbox{\boldmath$e$} ) = \frac{2
\pi}{\hbar} \vert M_{\nu' m'_s, \nu m_s} (\mbox{\boldmath$k$}) \vert^2
(f^0_{\nu m_s \mbox{\footnotesize{\boldmath$k$}} } - f^0_{\nu'm'_s
\mbox{\footnotesize{\boldmath$k$}}})\:
\delta \left( \varepsilon_{\nu' m'_s \mbox{\footnotesize{\boldmath$k$}} } -
\varepsilon_{\nu m_s \mbox{\footnotesize{\boldmath$k$}}} - \hbar \omega
\right),
\label{equ21}
\end{eqnarray}
where $M_{\nu' m'_s, \nu m_s}(\mbox{\boldmath$k$}) $ is the
inter-subband optical matrix element proportional to the amplitude
of the electromagnetic field $E_0$ and $f^0_{\nu m_s \mbox{\footnotesize{\boldmath$k$}}}$
is the distribution function in equilibrium. For the sake of
simplicity,  it is assumed  that the light intensity is low enough
to ignore a photoinduced redistribution of the symmetrical part
$f^{+}_{\nu m_s \mbox{\footnotesize{\boldmath$k$} }}$ = $( f_{\nu m_s
\mbox{\footnotesize{\boldmath$k$}} } + f_{\nu \bar{m_s}, - \mbox{\footnotesize{\boldmath$k$}} } )
/ 2$.

The most important result of the microscopic theory is that both
the initial and final states of the carriers involved in the
optical transition contribute to  the circular photogalvanic
current with different strength and directions. The partial
currents are proportional to the  group velocity being dependent
on~{\boldmath$k$}, the momentum relaxation time~$\tau_p$, and the
occupation of the initial states described by the distribution
function. Therefore the direction of total current depends of the
details of experimental conditions and may change its sign by
varying the radiation frequency, temperature etc. An interesting
feature of spin orientation induced CPGE at inter-band excitation
was pointed out in~\cite{Golub03p1,Golub03p2} showing that varying
the frequency around the fundamental band edge results in sign
inversion of the current due to  SIA contribution but not to BIA.

%
\begin{figure}[h]
\centerline{\epsfxsize 120mm \epsfbox{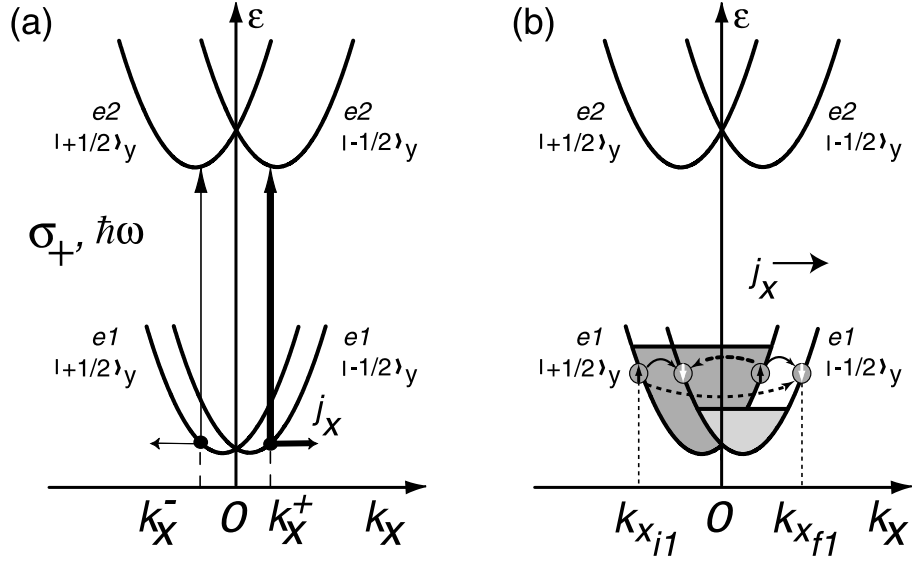  }}

\caption{Microscopic picture of (a) circular photogalvanic effect
and (b)  spin-galvanic effect at inter-subband excitation in
C$_{2v}$ point group samples. In (a) the current $j_x$ is caused
by the imbalance of optical transition probabilities at $k^-_x$
and $k^+_x$ decaying with the momentum relaxation time $\tau_p$.
Excitation with $\sigma_+$ radiation of $\hbar \omega$ less than
the energy subband separation at {\boldmath$k$}=0,
$\varepsilon_{21}$, induces direct spin-conserving transitions
(vertical arrows) at $k_x^-$ and $k_x^+$. The rates of these
transitions  are different as illustrated by the different
thickness of the arrows (reversing the angle of incidence mirrors
the  thicknesses of arrows).  This leads to a photocurrent due to
an asymmetrical distribution of carriers in {\boldmath $k$}-space
if the splitting of the $e${\it 1} and $e${\it 2} subbands is
non-equal. Increasing of the photon energy shifts more intensive
transitions to the left and less intensive to the right resulting
in a current sign change. In (b) the current occurs after
thermalization in the lowest subband which results in the spin
orientation in $e${\it 1} subband. This spin-galvanic current is
caused by asymmetric spin-flip scattering. The rate of spin-flip
scattering depends on the value of the initial and final
{\boldmath$k$}-vectors.  Thus transitions sketched by dashed
arrows yield an asymmetric occupation of both subbands and hence a
current flow which decays with the spin relaxation time $\tau_s$.
The magnitude of the spin polarization and hence the current
depends on the initial absorption strength but not on the momentum
{\boldmath$k$} of transition. Therefore the shape of the spectrum
of the spin-galvanic current  follows the absorption. }
\label{fig05}
\end{figure}

{\it Inter-subband transitions:} The microscopic theory of spin
orientation induced CPGE was also developed for direct
inter-subband transitions in $n$-type QWs for both $C_s$ and
$C_{2v}$ symmetry with the result that the current is proportional
to the derivative of the absorbance~\cite{PRB03inv}. For the
typical  condition of inter-subband transitions in $n$-type QWs
that the momentum relaxation time of photoexcited carriers in the
$e${\it 2} subband, $\tau_p^{(2)}$, is much less than that in the
$e${\it 1} subbands, $\tau_p^{(1)}$, it was obtained that for
$C_s$ symmetry
\begin{equation}
j_x \sim \left( \beta^{(2)}_{yx}+\beta^{(1) }_{yx}\right) \:
\frac{d \:\eta_{12}(\hbar \omega)}{d\: \hbar \omega} \:
\tau_p^{(1)} I \:P_{circ} \hat{e}_y,
\label{equ33}
\end{equation}
where  $\eta_{12}$ is the absorbance,  $I$ is the radiation
intensity, $\beta^{(1)}_{yx}$ and $\beta^{(2)}_{yx}$ are
components of {\boldmath$\beta$} in the $e${\it 1} and $e${\it 2}
subbands, respectively.  The analogue equation was obtained also
for QWs of $C_{2v}$ symmetry. In contrast to spin-flip processes
occurring for $C_s$ symmetry described above in $C_{2v}$ symmetry
due to selection rules the absorption of circularly polarized
radiation is spin-conserving~\cite{book}. The asymmetric
distribution of photo-excited electrons resulting in a current is
caused by these spin-conserving but spin-dependent transitions. It
was shown in~\cite{PRB03inv,IT} that under oblique excitation by
circularly polarized light the rates of inter-subband transitions
are different for electrons with the spin oriented co-parallel and
anti-parallel to the in-plane direction of light propagation. This
is depicted in Fig.~\ref{fig05}a by vertical arrows of different
thicknesses. In systems with {\boldmath$k$}-linear spin splitting
such processes lead to an asymmetrical distribution of carriers in
{\boldmath$k$}-space, i.e. to an electrical current. Also  for
these symmetry the photocurrent is proportional to the derivative
of absorption and is given by
\begin{equation}
j_x \sim \left( \beta^{(2)}_{yx}-\beta^{(1) }_{yx}\right) \:
\frac{d \:\eta_{12}(\hbar \omega)} {d\: \hbar \omega}
\:\tau_p^{(1)} I  \:P_{circ} \hat{e}_y.
\label{equ33a}
\end{equation}
 Since the circular photogalvanic effect in QW
structures of C$_{2v}$ symmetry is caused by spin-dependent {\it
spin-conserving} optical transitions, the photocurrent described
by Eq.~(\ref{equ33a}) in contrast to Eq.~(\ref{equ33}) is
proportional to the difference of subband spin splittings.

\subsubsection{One- and two-photon excitation}
\label{IIB4}

One more spin photocurrent
related to the photogalvanic effect was proposed in~\cite{Bhat00p5432}
and has most recently been observed  in bulk GaAs~\cite{Stevens02p4382}.
The  generation of the photocurrent is
 based on quantum interference of one- and two-photon
excitation~\cite{Hache97p306} which represents a coherent
photogalvanic effect~\cite{Entin89p664}. Irradiation of a
semiconductor sample with a coherent superposition of  laser beam
of frequency $\omega$ satisfying  $E_g/2 < \hbar\omega < E_g$  and
its second harmonic may yield an electric current if there are
definite phase relations of the radiation fields. Tuning of the
phase relation among the coherent beams allows a control of the
current. In order to obtain a spin photocurrent proper
polarization of both beams is required. If the radiation fields at
$\omega$ and $2\omega$ have either the same circular polarization
or orthogonal linear polarizations, quantum interference at
injection  distinguishes carriers of opposite spin which results
in a net spin flux. It is possible to obtain spin photocurrent
without electric current flow. The asymmetry of spin population in
{\boldmath$k$}-space appears as a consequence of spin-orbit
coupling. The photocurrent is mediated by a fourth rank tensor and
thus needs no symmetry restriction as in the case of one-photon
excitation photogalvanic effect. It may be present in materials
with a center of inversion. So far coherent control of a spin
photocurrent has been observed in  bulk semiconductors
(GaAs)~\cite{Stevens02p4382} but it may also be possible in QWs
and in asymmetric superlattices~\cite{Magarill01p652}.

\subsection{Spin-galvanic effect}
\label{IIC}

The picture of  spin photocurrents given so far involved the
asymmetry of the momentum distribution of photoexcited carriers,
i.e. the spin orientation induced CPGE. In addition to CPGE a spin
driven current may also occur even after momentum relaxation of
photoexcited carriers. It is due to an asymmetry of spin-flip
scattering of non-equilibrium  spin polarized carriers as shown in
Fig.~\ref{fig05}b. This current is caused by the spin-galvanic
effect and will be described in the following.

\subsubsection{Phenomenology}
\label{IIC1}

The spin-galvanic effect  is caused by spin relaxation of a
uniform non-equilibrium spin polarization in QWs of gyrotropic
symmetry~\cite{Nature02}. While this effect may occur at any
method of spin orientation e.g. electric spin injection, optical
spin polarization is also possible resulting in a spin
photocurrent. Phenomenologically, an electric current can be
linked to the electron's averaged spin polarization {\boldmath$S$}
by
\begin{equation} j_{\alpha} =
\sum_{\gamma}Q_{\alpha \gamma} S_{\gamma}
\label{equ22}.
\end{equation}
Like in the case of {\boldmath$k$}-linear terms and CPGE here we
have
 again a second rank pseudo-tensor {\boldmath$Q$} with the same
symmetry restrictions like {\boldmath$\beta$} and
{\boldmath$\gamma$}. Therefore in zinc-blende structure QW,
non-zero components of $Q_{\alpha\gamma}$ exist in contrast to the
corresponding bulk crystals~\cite{Ivchenko89p175,Ivchenko90p550}.
Due to tensor equivalence we have the same non-zero components of
the tensor {\boldmath$Q$} and their relations as discussed above
for {\boldmath$\beta$} and {\boldmath$\gamma$}.
 For $C_{2v}$ symmetry of
(001)-grown QWs only two linearly independent components, $Q_{xy}$
and $Q_{yx}$, may be non-zero so that
\begin{equation}
j_{x} = Q_{xy}S_y  \:,\: \: \: \: \: \: \: \: \: \: \: j_{y} =
Q_{yx}S_x\:. \label{equ23}
\end{equation}
Hence, a spin polarization driven current needs a spin component
lying in the plane of QWs. In $D_{2d}$ symmetry there is only one
independent tensor component $Q_{xy} = Q_{yx}$. In $C_s$ symmetry
of (113)-oriented QWs an additional tensor component
$Q_{xz^\prime}$ may be non-zero and the spin-galvanic current may
be driven by spins oriented normally to the plane of QW.

\subsubsection{Microscopic model}
\label{IIC2}

\begin{figure}
\centerline{\epsfxsize 130mm \epsfbox{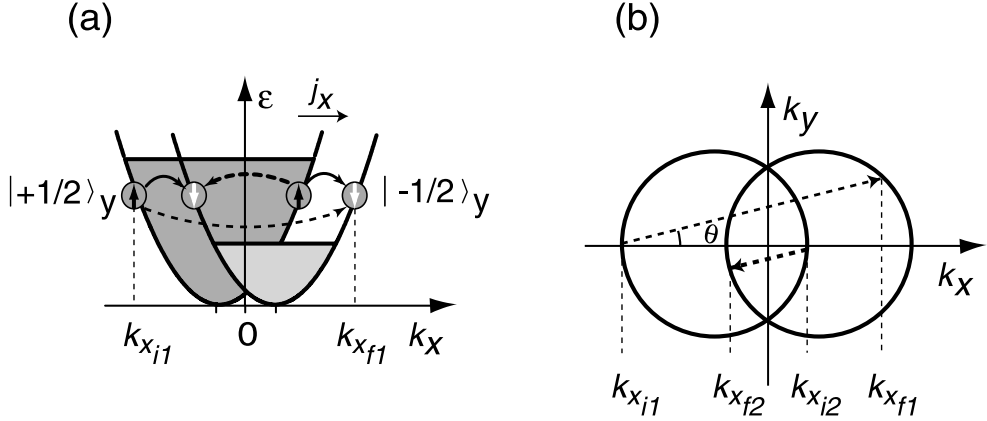  }}
\caption{Microscopic origin of the spin-galvanic current in the
presence of {\boldmath$k$}-linear terms in the electron
Hamiltonian. (a) one-dimensional sketch: the $\sigma_y k_x$ term
in the Hamiltonian splits the conduction band into two parabolas
with the spin $m_s=\pm 1/2$ in the $y$-direction. If one spin
subband is preferentially occupied, e.g., by spin injection (the
($|+1/2\rangle_y$ -states in the figure) asymmetric spin-flip
scattering results in a current in $x$-direction. The rate of
spin-flip scattering depends on the value of the initial and final
{\boldmath$k$}-vectors.  Thus transitions sketched by dashed
arrows yield an asymmetric occupation of both subbands and hence a
current flow. These transitions are also shown in two dimensions
(b) by dashed arrows at scattering angle $\theta$. If instead of
the spin-down subband the spin-up subband is preferentially
occupied the current direction is reversed. }
\label{fig06}
\end{figure}

Microscopically, the spin-galvanic effect is caused by asymmetric
spin-flip relaxation of  spin polarized  electrons in systems with
{\boldmath$k$}-linear contributions to the effective
Hamiltonian~\cite{Nature02}. Fig.~\ref{fig06}a sketches the
electron energy spectrum along $k_x$ with the spin dependent term
$\beta_{yx}\sigma_y k_x$ resulting from BIA and SIA. In  this case
$\sigma_y$ is a good quantum number. Spin orientation in
$y$-direction causes the unbalanced population in spin-down and
spin-up subbands. The current flow is caused by
{\boldmath$k$}-dependent spin-flip relaxation processes. Spins
oriented in $y$-direction are scattered along $k_x$ from the
higher filled, e.g.   spin subband $|+1/2 \rangle_y$ to the less
filled  spin subband $|-1/2 \rangle_y$. Four quantitatively
different spin-flip scattering events exist and are sketched in
Fig.~\ref{fig06}a  by bent arrows. The spin-flip scattering rate
depends on the values of the wavevectors of the initial and the
final states~\cite{Averkiev02pR271}. Therefore spin-flip
transitions, shown by solid arrows in Fig.~\ref{fig06}a, have  the
same rates. They preserve the symmetric distribution of carriers
in the subbands and, thus, do not yield a  current. However, the
two scattering processes shown by broken arrows are inequivalent
and generate an asymmetric carrier distribution around the subband
minima in both subbands. This asymmetric population results in a
current flow along the $x$-direction. Within this model of elastic
scattering the current is not spin polarized since the same number
of spin-up and spin-down electrons move in the same direction with
the same velocity.

%

It must be pointed out that the above one-dimensional model, which
in a clear way demonstrates how a spin-galvanic  current can
occur, somehow simplifies the microscopic picture. The probability
of the spin-flip processes $| + 1/2, \mbox{\boldmath$k$}_i
\rangle_y$ $\rightarrow$ $| - 1/2 , \mbox{\boldmath$k$}_f
\rangle_y$ shown by arrows in Fig.~\ref{fig06} is given by the
product $[v(\mbox{\boldmath$k$}_i - \mbox{\boldmath$k$}_f)]^2
(\mbox{\boldmath$k$}_f + \mbox{\boldmath$k$}_i)^2$~(see Eq.~(30)
of \cite{Averkiev02pR271}). The amplitude $v(\mbox{\boldmath$k$}_f
- \mbox{\boldmath$k$}_i)$ depends on $\mbox{\boldmath$k$}_f -
\mbox{\boldmath$k$}_i$ and therefore the spin-flip process is
asymmetric as needed for the occurrence of the current. However,
for the one-dimensional model presented above the probability is
given by  $[v(k_{x_f} - k_{x_i})]^2 (k_{x_f} + k_{x_i})^2$. In the
case of  elastic scattering, as sketched in Fig.~\ref{fig06}a, the
magnitudes of the initial and final wavevectors are equal,
$|k_{x_i}|=|k_{x_f}|$, thus $k_{x_f} + k_{x_i} =0$ and the
probability vanishes. A non-zero current is obtained at inelastic
scattering and at elastic scattering with $k_y \neq 0$. The latter
situation is depicted in Fig.~\ref{fig06}b.

Note that the reverse process to the spin-galvanic effect i.e. a
spin polarization induced by an electric current flow has been
theoretically considered in~\cite{Aronov89p431,Edelstein89p233}.

The uniformity of  spin polarization in space is preserved during
the scattering processes. Therefore the spin-galvanic effect
differs  from other experiments  where the spin current is caused
by inhomogeneities. These effects will be discussed in section~6.

\subsubsection{Microscopic theory}
\label{IIC3}

The microscopic theory of the spin-galvanic effect has been
developed in~\cite{PRB03sge}
 for inter-subband
transitions in $n$-type zinc-blende structure materials of
$C_{2v}$ symmetry. In this case the spin orientation (see
Fig.~\ref{fig05}b) is generated by resonant spin-selective optical
excitation (see Fig.~\ref{fig05}a) followed by spin-non-specific
thermalization.

The occurrence of a current  is due to the spin dependence of the
electron scattering matrix elements $M_{
\mbox{\footnotesize{\boldmath$k$}}^\prime
\mbox{\footnotesize{\boldmath$k$}}}$. The   2 $\times$ 2 matrix
$\hat{M}_{ \mbox{\footnotesize{\boldmath$k$}}^\prime
\mbox{\footnotesize{\boldmath$k$}}  }$ can be written as a linear
combination of the unit matrix $\hat{I}$ and Pauli matrices as
follows
\begin{equation}
\hat{M}_{ \mbox{\footnotesize{\boldmath$k$}}^\prime
\mbox{\footnotesize{\boldmath$k$}}  } = A_{
\mbox{\footnotesize{\boldmath$k$}}^\prime
\mbox{\footnotesize{\boldmath$k$}}  } \hat{I} + \mbox{\boldmath$
\sigma$} \cdot \mbox{\boldmath$ B$}_{
\mbox{\footnotesize{\boldmath$k$}}^\prime
\mbox{\footnotesize{\boldmath$k$}} } \:,
 \label{equ24}
 \end{equation}
where $A^*_{ \mbox{\footnotesize{\boldmath$k$}}^\prime
\mbox{\footnotesize{\boldmath$k$}} } =A_{
\mbox{\footnotesize{\boldmath$k$}}
\mbox{\footnotesize{\boldmath$k$}}^\prime}$, $B^*_{
\mbox{\footnotesize{\boldmath$k$}}^\prime
\mbox{\footnotesize{\boldmath$k$}}  } = B_{
\mbox{\footnotesize{\boldmath$k$}}
\mbox{\footnotesize{\boldmath$k$}}^\prime}$ due to hermiticity of
the interaction and $A_{-
\mbox{\footnotesize{\boldmath$k$}}^\prime, -
\mbox{\footnotesize{\boldmath$k$}}  } =A_{
\mbox{\footnotesize{\boldmath$k$}}
\mbox{\footnotesize{\boldmath$k$}}^\prime}$, $B_{-
\mbox{\footnotesize{\boldmath$k$}}^\prime, -
\mbox{\footnotesize{\boldmath$k$}}  } = - B_{
\mbox{\footnotesize{\boldmath$k$}}
\mbox{\footnotesize{\boldmath$k$}}^\prime}$ due to the symmetry
under time inversion. The spin-dependent part of the scattering
amplitude in (001)-grown QW structures is given
by~\cite{Averkiev02pR271}
\begin{equation}
\mbox{\boldmath$ \sigma$} \cdot  \mbox{\boldmath$ B$}_{
\mbox{\footnotesize{\boldmath$k$}}^\prime
\mbox{\footnotesize{\boldmath$k$}}  } = v( \mbox{\boldmath$k$}  -
\mbox{\boldmath$k$}^\prime) [ \sigma_x (k^\prime_y + k_y) -
\sigma_y (k^\prime_x + k_x)] \:.
\label{equ25}
\end{equation}
We note that Eq.~(\ref{equ25}) determines the spin relaxation
time, $\tau^\prime_s$, due to the Elliot-Yafet mechanism. The
spin-galvanic current, for instance  in $y$ direction, has the
form~\cite{PRB03sge}
\begin{equation} \label{equ25a}
j_{SGE,x}= Q_{xy}S_y \sim e\: n_e \frac{\beta_{yx}^{(1)}}{\hbar}
\frac{\tau_p }{\tau^\prime_s} S_y\:,\: \: \: \: \: \: \: \: \: \:
\:j_{SGE,y}= Q_{yx}S_x \sim e\: n_e \frac{\beta_{xy}^{(1)}}{\hbar}
\frac{\tau_p }{\tau^\prime_s} S_x \:.
\end{equation}
Since scattering is the origin of the spin-galvanic effect,  the
spin-galvanic current, $j_{SGE}$, is determined by the
Elliot-Yafet spin relaxation time. The relaxation time
$\tau^\prime_s$ is proportional to the momentum relaxation time
$\tau_p$.  Therefore the ratio $\tau_p / \tau_s^\prime$ in
Eqs.~(\ref{equ25a}) does not depend on the momentum relaxation
time. The in-plane average spin $S_x$ in Eqs.~(\ref{equ25a})
decays with the total spin relaxation time $\tau_s$ (which may
have a contribution from any spin relaxing process). Thus the time
decay of the spin-galvanic current following a pulsed
photoexcitation is determined by $\tau_s$.

For the case, where spin relaxation is obtained as a result of
inter-subband absorption of circularly polarized radiation, the
current is given by
\begin{equation}
j_{SGE,x} = Q_{xy}S_y \sim e\:\frac{\beta_{yx}^{(1)}}{\hbar}
\frac{\tau_p \tau_s}{\tau^\prime_s} \frac{\eta_{21}I}{\hbar
\omega}
 P_{circ} \xi \hat{e}_y \:,\: \: \: \: \: \: \: \: \: \: \:
j_{SGE,y} = Q_{yx}S_x \sim e\:\frac{\beta_{xy}^{(1)}}{\hbar}
\frac{\tau_p \tau_s}{\tau^\prime_s} \frac{\eta_{21}I}{\hbar
\omega}
 P_{circ} \xi \hat{e}_x \:. \label{equ29}
\end{equation}
where $\eta_{21}$ is the
 absorbance at the transitions between $e${\it 1} and $e${\it 2} subbands.
 The parameter $\xi$, varying between 0 and 1,                                                                                                                                                                                                                                                                          is the ratio of
photoexcited electrons relaxing to the $e1$ subband with and
without spin-flip. It determines the degree of spin polarization
in the lowest subband (see Fig.~\ref{fig04}b) and depends on the
details of the relaxation mechanism. Optical orientation requires
$\xi\neq 0$~\cite{Meier,IT,Parson71p1850}. Eqs.~(\ref{equ29}) show
that the  spin-galvanic current is proportional to the absorbance
and is determined by the spin splitting in the first subband,
$\beta^{(1)}_{yx}$ or $\beta^{(1)}_{xy}$.

\subsubsection{Spin-galvanic effect at optical orientation}
\label{IIC4}

Excitation of QWs by circularly polarized light results in a spin
polarization which, at proper orientation of the electron spins,
causes a photocurrent due to the spin-galvanic effect. Because of
the tensor equivalence the spin-galvanic current induced by
circularly polarized light always occurs  simultaneously with the
spin orientation induced CPGE. The two effects  can be separated
by time-resolved measurements because of the different relaxation
mechanisms of the two currents. After removal of light  or under
pulsed photoexcitation the circular photogalvanic current decays
with the momentum relaxation time whereas the spin-galvanic
current decays with the spin relaxation time. On the other hand,
as it has been recently shown~\cite{PRB03sge}, at inter-subband
transitions the spin-galvanic effect may be separated from CPGE
making use of the spectral behaviour at resonance. The optically
induced spin-galvanic current reproduces  the absorbance whereas
CPGE is proportional to the derivative of the absorbance and
vanishes at the resonance frequency. This will be discussed in
more detail in section~4.2.2.

\subsubsection{Spin-galvanic effect at optical orientation in the presence of magnetic field}
\label{IIC5}

\begin{figure}
\centerline{\epsfxsize 46mm \epsfbox{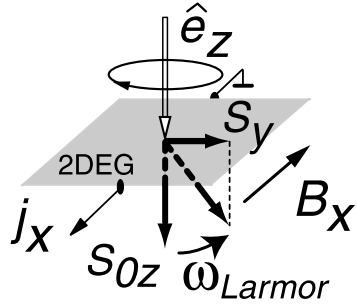  }}
\caption{Optical scheme of generating a uniform  in-plane
spin-polarization which causes spin-galvanic photocurrent.
Electron spins are oriented normal to the plane of QW by
circularly polarized radiation and rotated into the plane by
Larmor precession in a magnetic field $B_x$. }
\label{fig07}
\end{figure}

Another possibility to investigate the spin-galvanic effect
without contributions of the spin orientation induced CPGE to the
current has been introduced in~\cite{Nature02}. This method is not
limited to resonant inter-subband optical excitation. The spin
polarization was  obtained  by absorption of circularly polarized
radiation at normal incidence on (001)-grown QWs as depicted in
Fig.~\ref{fig07}. For normal incidence the spin orientation
induced CPGE as well as the spin-galvanic effect vanish because
$\hat{e}_x=\hat{e}_y=0$~(see Eqs.~(\ref{equ14}))  and $S_x=S_y=0$
(see Eqs.(\ref{equ23})), respectively. Thus, we obtain a spin
orientation along the $z$ coordinate but no spin photocurrent.

The steady-state spin polarization $S_{0z}$ is proportional to the
spin generation rate $\dot{S}_z$. To obtain an in-plane component
of the spins, necessary for the spin-galvanic effect, a magnetic
field {\boldmath$B$}$\parallel x$ has been  applied. Due to Larmor
precession a non-equilibrium spin polarization $S_y$ is induced
being
\begin{equation}
S_y = - \frac{\omega_L \tau_{s\perp} }{1 + (\omega_L \tau_s )^2}\:
S_{0z}\:,
\label{equ30}
\end{equation}
where $\tau_s = \sqrt{\tau_{s\parallel} \tau_{s\perp} }$,
$\tau_{s\parallel}$, $\tau_{s\perp}$ are the longitudinal and
transverse electron spin relaxation  times, $\omega_L$ is the
Larmor frequency.
 The denominator in Eq.~(\ref{equ30}) yielding the
decay of $S_y$ for $\omega_L$ exceeding the inverse spin
relaxation time is well known from the Hanle
effect~\cite{Meier,Hanle24p93}

%

On a phenomenological level and at small magnetic fields,
$\omega_L\tau_s\ll 1$, this magnetic field induced spin
photocurrent  can be described  by
\begin{equation}
j_{\alpha} = \sum_{\beta\gamma}\mu_{\alpha \beta \gamma}
B_{\beta}\: i \left( {\boldmath E} \times {\boldmath E}^*
\right)_{\gamma} = \sum_{\beta\gamma}\mu_{\alpha \beta \gamma}
B_{\beta} \hat{e}_{\gamma} E^2_0 P_{circ}\;,
\label{equ31}
\end{equation}
where $\mu_{\alpha\beta\gamma}$ is a third rank tensor. As
$P_{circ}$ is a pseudo-scalar and {\boldmath$B$} a pseudo-vector,
$\mu_{\alpha\beta\gamma}$ is a regular negative-parity third rank
tensor which is allowed in inversion asymmetric materials only.
Gyrotropy at zero magnetic field, as in the case of only optical
excited spin-galvanic effect or of circular photogalvanic effect,
is not necessary. We note that in non-gyrotropic $p$-type bulk
GaAs a magnetic field induced circular photogalvanic effect was
previously observed at intra-band
excitation~\cite{Andrianov84p882}. However, this effect is not due
to spin orientation  and does not occur in $p$-type QWs due to
spatial quantization~\cite{Ivchenko89p663}. In QWs under normal
incidence of the light and for a magnetic field lying in the plane
of a QW of $C_{2v}$ symmetry, which corresponds to the
measurements in section~4.2.1 the current is described by two
independent components of the tensor {\boldmath$\mu$} and can be
written as
\begin{equation}
j_{x}= \mu_{xxz}B_{x} \hat{e}_{z} E^2_0 P_{circ} \:,\: \: \: \: \:
\: \: \: \: \: \:j_{y}= \mu_{yyz}B_{y} \hat{e}_{z}E^2_0 P_{circ}
\:. \label{equ32}
\end{equation}
The current  {\boldmath$j$} and the magnetic field {\boldmath$B$}
are parallel (or anti-parallel) when the magnetic field is applied
along $\langle 110\rangle$ and neither parallel nor perpendicular
for {\boldmath$B$}$\parallel \langle 100\rangle$. In
D$_{2d}$-symmetry QWs with symmetric interfaces $\mu_{xxz} =
-\mu_{yyz}$ and therefore the current is perpendicular to the
magnetic field for {\boldmath$B$}$\parallel \langle 100\rangle$.

\subsection{Spin orientation induced circular photogalvanic effect versus spin-galvanic effect}
\label{IID}

The spin orientation induced circular photogalvanic effect and the
spin-galvanic effect have in common that the current flow is
driven by an asymmetric distribution of carriers in
{\boldmath$k$}-space in systems with lifted spin degeneracy due to
{\boldmath$k$}-linear terms in the Hamiltonian.  The crucial
difference between both effects is, that the spin-galvanic effect
may be caused by any means of spin injection, while the spin
orientation induced CPGE needs optical excitation with circularly
polarized radiation. Even if the spin-galvanic effect is achieved
by optical spin orientation, as discussed here, the microscopic
mechanisms are different. The spin-galvanic effect is caused by
asymmetric spin-flip scattering of spin polarized carriers and it
is determined by the process of spin relaxation (see
Fig.~\ref{fig06}). If spin relaxation is absent, the spin-galvanic
current  vanishes. In contrast, the spin orientation induced CPGE
is  the result of selective  photoexcitation of carriers in
{\boldmath$k$}-space with circularly polarized light due to
optical selection rules and depends on momentum relaxation (see
Fig.~\ref{fig02}).  In some optical experiments the observed
photocurrent may represent a sum of both effects. For example, if
we irradiate an (001)-oriented QW at oblique incidence of
circularly polarized radiation, we obtain both, selective
photoexcitation of carriers in {\boldmath$k$}-space determined by
momentum relaxation and spin-galvanic effect due to an in-plane
component of non-equilibrium spin polarization. Thus both effects
contribute to the current occurring in the plane of the QW. The
two mechanisms can be distinguished by time resolved measurements.

\section{Methods }
\label{III}

\subsection{ Samples}
\label{IIIA}

The experiments were  carried out on GaAs, InAs, semimagnetic
BeZnMnSe  and SiGe heterostructures belonging to two different
classes of symmetry. Higher symmetric structures were
(001)-oriented QWs. While these structures can belong to two point
groups, either $D_{2d}$ or $C_{2v}$ our measurements showed that
all samples available for the present work correspond  to the
point group $C_{2v}$. Structures of the lower symmetry class $C_s$
were (113)-oriented QWs and quantum wells grown on (001)-miscut
substrates.

Zinc-blende structure based QW samples were molecular-beam-epitaxy
(MBE) grown $n$- and $p$-type GaAs/AlGaAs with QW widths $L_W$ of
4~nm to 20~nm~\cite{PRL01,PRB03inv,PRL02}, $n$-type InAs/AlGaSb
QWs with  $L_W$=15~nm ~\cite{PRL01,Behet98p428}  as well as single
$n$-type GaAs heterojunctions. Free carrier densities, $n_s$ for
electrons and $p_s$ for holes, ranged from $10^{11}$~cm$^{-2}$ to
$2\cdot 10^{12}$~cm$^{-2}$. The mobility at 4.2~K in $n$-type
samples was from $5\cdot 10^5$~cm$^2$/Vs to $2
\cdot10^6$~cm$^2$/Vs and in $p$-type samples was about $5\cdot
10^5$~cm$^2$/Vs.

Semimagnetic BeZnMnSe semiconductor heterostructures were grown by
MBE on semi-insulating GaAs substrates with (001)
orientation~\cite{DPG01,Fiederling99p787}. The heterostructures
consisted of a 500-nm-thick Be$_{0.03}$Zn$_{0.97}$Se layer
$n$-doped to $2 \cdot 10^{18}$ cm$^{-3}$ followed by an 100~nm
thick Be$_{0.05}$Zn$_{0.89}$Mn$_{0.06}$Se layer $n$-doped to $6
\cdot 10^{18}$ cm$^{-3}$.

%
\begin{figure}
\centerline{\epsfxsize 86mm \epsfbox{ 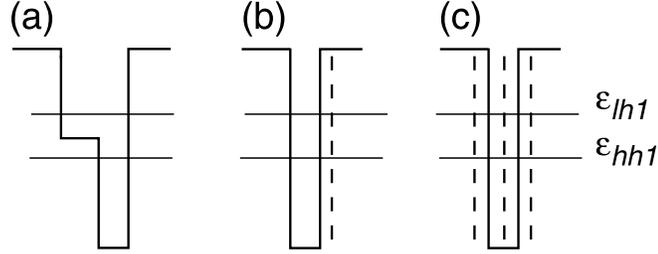 }}
\caption{Potential profiles of SiGe QW samples: (a)
compositionally stepped QW, (b) asymmetrically  doped
compositionally symmetric QW, and (c) symmetric QW. The vertical
dashed lines indicate the doping. }
\label{fig08}
\end{figure}

The measurements on SiGe QWs were carried out on $p$-type
structures MBE-grown on (001)- and (113)-oriented Si
substrates~\cite{PRB02}. Both bulk Si as well as Ge have a center
of inversion, therefore in order to obtain gyrotropy asymmetric
QWs were grown. Two groups of (001)-grown  asymmetric samples,
whose potential profiles are sketched in Figs.~\ref{fig08}a
and~\ref{fig08}b, were fabricated in the following manner: one of
the groups of samples was compositionally stepped
(Fig.~\ref{fig08}a) comprising 10 QWs
[Si$_{0.75}$Ge$_{0.25}$(4~nm)/ Si$_{0.55}$Ge$_{0.45}$(2.4~nm)]
separated by 6~nm Si barriers. The second group of asymmetric
structures had a single QW of Si$_{0.75}$Ge$_{0.25}$ composition
which was doped with boron from one side only (Fig.~\ref{fig08}b).
These structures are of the $C_{2v}$ point group symmetry which
was  confirmed by the experiments described below. Structures of
the lower symmetry $C_s$ were (113)-grown with a
Si/Si$_{0.75}$Ge$_{0.25}$(5~nm)/Si single QW one-side boron doped.
As a reference sample a (001)-grown compositionally symmetric and
symmetrically boron doped multiple QW structure
(Fig.~\ref{fig08}c) of sixty Si$_{0.7}$Ge$_{0.3}$(3~nm) QW has
been used. All these samples had free carrier densities $p_s$ of
about $8 \cdot 10^{11}$~cm$^{-2}$ in each QW.

%
\begin{figure}
\centerline{\epsfxsize 100mm \epsfbox{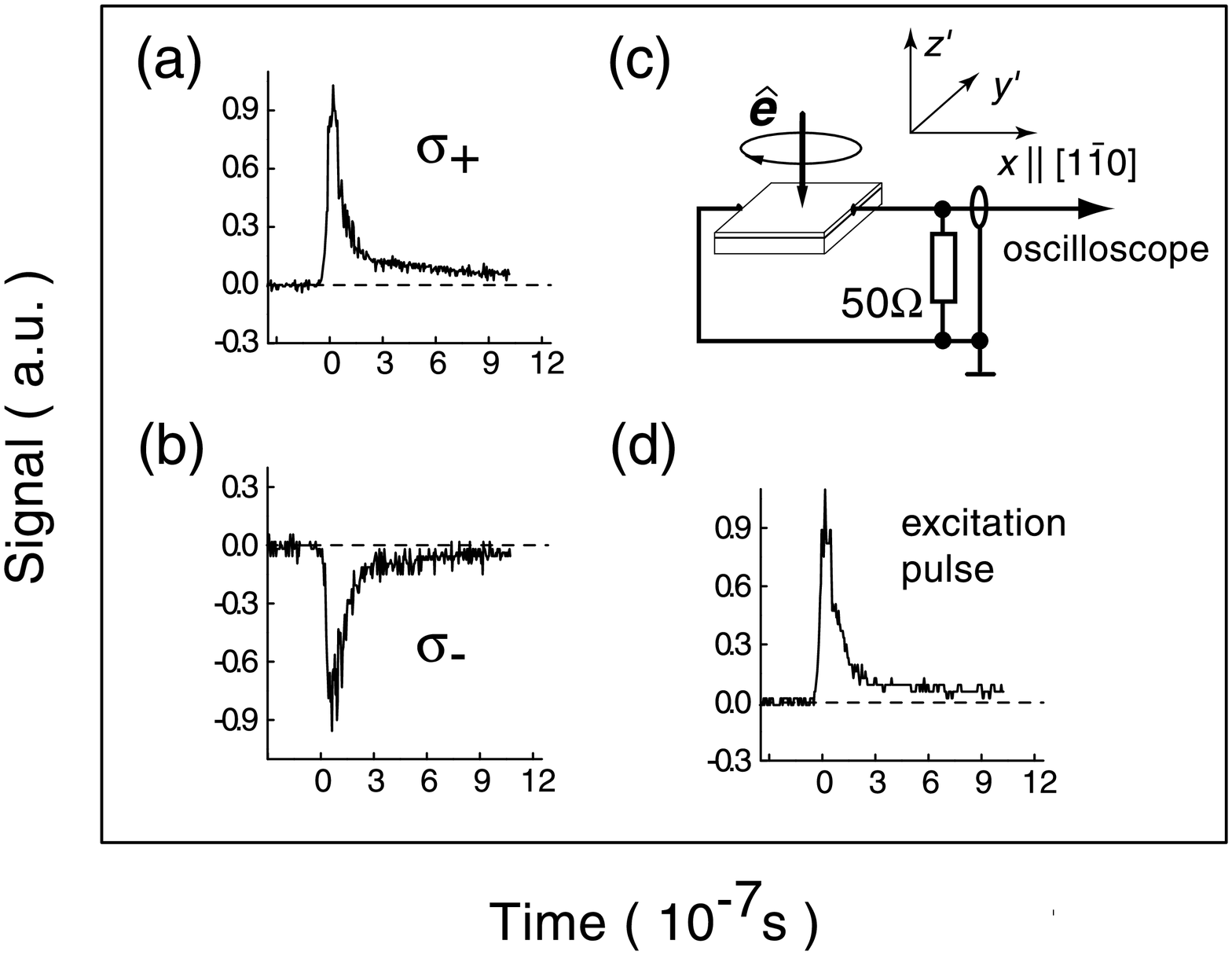}}
\caption{Oscilloscope traces obtained for pulsed excitation of
(113)-grown $n$-type GaAs QWs at $\lambda = 10.6~\mu$m. (a) and
(b) show CPGE signals obtained for $\sigma_{+}$ and
$\sigma_{-}$-circular polarization, respectively. For comparison
in (d) a signal pulse of a fast photon drag detector is plotted.
In (c) the measurement arrangement is sketched. For (113)-grown
samples being of C$_s$ symmetry  radiation was applied at normal
incidence and the current detected in the direction $x \parallel$
$[1\bar{1}0]$. For (001)-grown QWs oblique incidence was used in
order to obtain the helicity dependent current.}
\label{fig09}
\end{figure}

\begin{figure}[h]
\centerline{\epsfxsize 86mm \epsfbox{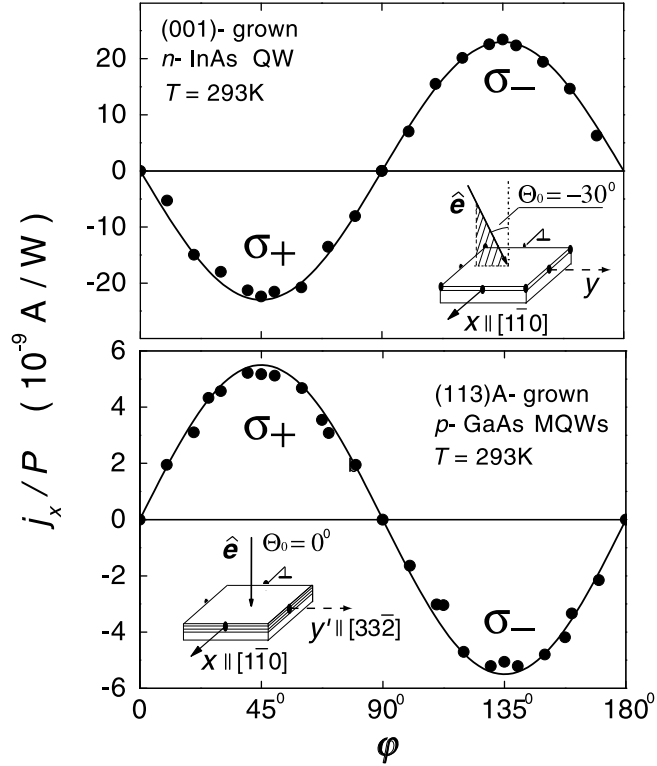  }}
\caption{Photocurrent  in QWs normalized by the light power $P$ as
a function of the phase angle $\varphi$ defining helicity.
Measurements are presented for $T=~$293~K and $\lambda=76~\mu$m.
The  insets  show   the geometry of the experiment. Upper panel:
oblique incidence of radiation with an  angle of incidence
$\Theta_0 = -30^\circ$ on $n$-type (001)- grown InAs/AlGaSb QWs
(symmetry class $C_{2v}$). The current $j_{x }$ is perpendicular
to the direction of light propagation. Lower panel: normal
incidence of radiation on $p$-type (113)A- grown GaAs/AlGaAs QWs
(symmetry class C$_s$). The current $j_{x }$ flows along
[1$\bar{1}$0]- direction perpendicular to the mirror plane of
$C_s$ symmetry. Full lines show ordinate scale fits after
Eqs.~(\protect \ref{equ14}) and~\protect (\ref{equ15}) for the top
and lower panel, respectively.}
\label{fig10}
\end{figure}

The sample edges were oriented along the [1$\bar{1}$0]- and
[33$\bar{2}$]- directions for the (113)-grown sample and along the
[1$\bar{1}$0]- and [110]- directions for the (001)-grown. For
(113)-oriented samples two pairs of ohmic contacts   were centered
along  opposite sample edges pointing in the directions
$x\parallel $ [1$\bar{1}$0] and $y \parallel $ [33$\bar{2}$] (see
Fig.~\ref{fig09} and  inset in Fig.~\ref{fig10}, lower plate). For
(001)-oriented samples two pairs of  point  contacts in the middle
of the sample edges with connecting lines along $x
\parallel$ [1$\bar{1}$0] and $y^\prime \parallel$ [110] were   prepared
(see inset in Fig.~\ref{fig10}, upper plate). These samples had
two additional pairs of contacts at the corners of the samples
corresponding to the $\langle 100 \rangle$-directions (see inset
in Fig.~\ref{fig10}, upper plate).

\subsection{Experimental technique}
\label{IIIB}

For  optical excitation mid-infrared (MIR), far-infrared (FIR) and
visible laser radiation was used. Most of the measurements were
carried out in the infrared with photon energies less than the
energy gap of  investigated semiconductors. For investigations of
spin photocurrents infrared excitation has several advantages.
First of all below the energy gap the absorption is very weak and
therefore allows homogeneous excitation with marginal heating of
the 2D electron gas. Furthermore, in contrast to inter-band
excitation, there are no spurious photocurents due to other
mechanisms like the Dember effect, photovoltaic effects at
contacts and Schottky barriers etc. Depending on the photon energy
and QW band structure the MIR and FIR radiation induce direct
optical transitions between size quantized subbands in $n$- and
$p$-type samples or, at longer wavelength, indirect  optical
transitions (Drude absorption) in the lowest subband.

A high power pulsed mid-infrared (MIR) transversely excited
atmospheric pressure carbondioxide (TEA-CO$_2$) laser and a
molecular far-infrared (FIR) laser~\cite{PhysicaB99tun,JPC02tun}
have been used  as radiation sources in the spectral range between
9.2~$\mu$m and 496~$\mu$m. The corresponding photon energies
$\hbar \omega$  lie in the range of 135~meV to 2~meV. The
radiation pulses ($ \simeq $100~ns) of a power  $P$ up to 50~kW
were focused to a spot of about 1~mm$^2$ yielding a maximum
intensity  of about 5~MW/cm$^2$. Such high intensities are only
needed for saturation measurements describes in section~4.4. The
power required to detect spin photocurrents is much lower. One
series of measurements was carried out making use of  the
frequency tunability of the free electron laser ''FELIX'' at
FOM-Rijnhuizen in The Netherlands~\cite{Knippels99p1578}. The
FELIX operated in the spectral range between 7~$\mu$m and
12~$\mu$m. The output pulses of light from FELIX were chosen to be
3 ps long, separated by 40~ns, in a train (or ''macropulse'') of
5~$\mu$s duration. The macropulses had a repetition rate of 5~Hz.

Typically these lasers emit linearly polarized radiation. The
polarization  was modified from linear to circular using a Fresnel
rhomb and $\lambda/4$ plates for MIR and FIR radiation,
respectively. The helicity $P_{circ}$ of the incident light was
varied from $-1$ (left-handed circular, $\sigma_-$) to $+1$
(right-handed circular, $\sigma_+$) according to $P_{circ} =
\sin{2 \varphi}$ (see Eq.~(\ref{equ12})). In the present
experimental arrangement the phase angle $\varphi$ corresponds to
the angle between the initial plane of polarization and  the
optical axis of the $\lambda/4$ plate or the polarization plane of
the Fresnel rhomb.

For optical inter-band excitation a cw-Ti:Sapphire laser was used
providing radiation of $\lambda$= 0.777~$\mu$m with about 100~mW
power. In order to extract the helicity dependent current the
linearly polarized laser beam was transmitted through a
photoelastic modulator which yields  a periodically oscillating
polarization between $\sigma_+$ and $\sigma_-$~\cite{Nature02}.

Samples were studied at room temperature or mounted in an optical
cryostat which allowed the variation of temperature in the  range
of 4.2~K to 293~K. The  photocurrent $j_x$ was measured in the
unbiased structures via the voltage drop across a 50~$\Omega$ load
resistor in a closed circuit configuration~\cite{APL00} (see
Fig.~\ref{fig09}c). The current in the case of the excitation by
visible radiation was recorded by a lock-in amplifier in phase
with the photoelastic modulator.

The experiments on the spin-galvanic effect, which require an
external magnetic field, were performed at room temperature in a
conventional electromagnet with the magnetic field up to 1~T and
at 4.2~K using a superconducting split-coil magnet with $B$ up to
3~T.

%

\section{Experimental results and discussion}
\label{IV}

\subsection{Spin polarization induced circular photogalvanic effect}
\label{IVA}

\subsubsection{General features}
\label{IVA1}

With illumination of  QW structures by polarized radiation a
current signal proportional to the helicity $P_{circ}$ has been
observed in unbiased samples~\cite{APL00,PRL01,PRB02,DPG01}. The
irradiated QW structure represents a current source wherein the
current flows in the QW. Fig.~\ref{fig09} shows measurements of
the voltage drop across a 50~$\Omega$ load resistor  in response
to 100~ns laser pulses at $\lambda = 10.6~\mu$m. Signal traces are
plotted in Fig.~\ref{fig09}a for right-  and in Fig.~\ref{fig09}b
for left-handed  circular polarization in comparison to a
reference signal (see Fig.~\ref{fig09}d) obtained from the fast
photon-drag detector~\cite{Ganichev84p20,Beregulin90p853}. The
signal follows the temporal structure of the applied laser pulses.
In Fig.~\ref{fig10} the current is shown as a function of the
phase angle $\varphi$. The current signal assumes a maximum at
circular polarized radiation  and changes sign if the polarization
is switched from $\sigma_+$ to $\sigma_-$. In the case of linearly
polarized radiation corresponding to $\varphi = 0^{\circ}$ or
$90^{\circ}$ the current vanishes. The radiation induced current
and its characteristic helicity dependence reveals that we are
dealing with the circular photogalvanic effect. The effect is
quite general and has been observed  in all samples in the
temperature range of 4.2~K to 293~K and in a wide spectral range.

In (001)-oriented samples a helicity dependent signal is only
observed under oblique incidence~\cite{PRL01,PhysicaE02,PRB02}.
For light propagating along $\langle 110 \rangle$ direction the
photocurrent flows perpendicular to the wavevector of the incident
light (see Fig.~\ref{fig10}, upper plate). This observation is in
accordance to Eqs.~(\ref{equ14}).  For illumination along a cubic
axis $\langle 100 \rangle$  both a transverse and longitudinal
circular photogalvanic current are detected~\cite{ICPS26invited}.
The presence of a transverse current in this geometric
configuration in all  (001)-oriented samples, except miscut,
investigated as yet unambiguously demonstrates that they belong to
the symmetry class $C_{2v}$. Indeed, in such a geometry
($\hat{e}_x = \hat{e}_y = 1/\sqrt{2}$), the  transverse effect is
only allowed for the $C_{2v}$ symmetry class and is forbidden for
$D_{2d}$  symmetry as it can be seen from Eqs.~(\ref{equ13}) and
(\ref{equ14}) and the discussion following. In samples grown on a
(113)-GaAs surface or on (001)-miscut substrates representing the
lower symmetry class $C_s$,  the CPGE has been  observed also
under normal incidence of
radiation~\cite{APL00,PRL01,PhysicaE02,PRB02}  as shown in the
lower plate of Fig.~\ref{fig10}. This is  in contrast to
(001)-oriented samples and in accordance to the phenomenological
theory of the CPGE for $C_s$ (see Eqs.~(\ref{equ15})). For normal
incidence in this symmetry  the current always flows along the
[1$\bar{1}$0]- direction perpendicular to the plane of mirror
reflection of the point group $C_s$. The solid lines in
Fig.~\ref{fig10} are obtained from the  phenomenological picture
outlined  above  which perfectly describes the experimental
observations.

%
\begin{figure}[h]
\centerline{\epsfxsize 86mm \epsfbox{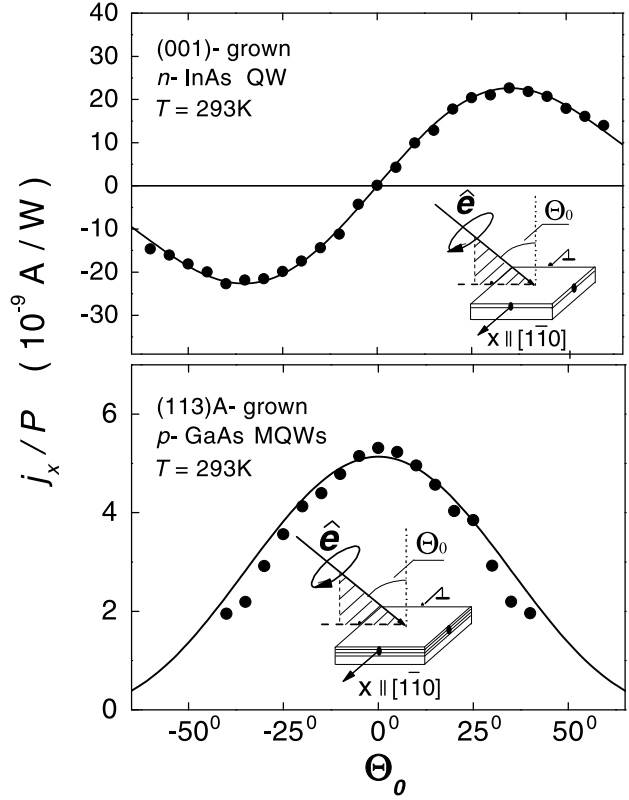  }}
\caption{Photocurrent in QWs   normalized by $P$ as a function of
the angle of incidence $\Theta_0$ for right-handed circularly
polarized radiation $\sigma_+$ measured perpendicular to light
propagation ($T=$~293~K, $\lambda=76~\mu$m). Upper panel: $n$-type
(001)-grown InAs/AlGaSb QWs (C$_{2v}$). Lower panel: $p$-type
(113)A-grown  GaAs/AlGaAs QWs (C$_s$). Full lines show ordinate
scale fits after Eqs.~(\protect \ref{equ14})~(\protect
\ref{equ16}) (upper plate) and~\protect (\ref{equ15})~(\protect
\ref{equ17}) (lower plate).}
\label{fig11}
\end{figure}

In Fig.~\ref{fig11} we show the dependence of the photocurrent on
the angle of incidence $\Theta_0$ of the right-handed circularly
polarized laser beam. For (001)-oriented samples
($C_{2v}$-symmetry) a variation of $\Theta_0$ in the plane of
incidence normal to $x$ changes the sign of the current $j_x$ at
normal incidence, $\Theta_0$=0, as can be seen in the upper panel
of Fig.~\ref{fig11}. The lower panel of Fig.~\ref{fig11} displays
the angular dependence for (113)-oriented quantum wells
($C_s$-symmetry). The  currents measured as a function of the
angle of incidence $\Theta_0$ along any direction in the plane of
(001)-oriented samples and along $x
\parallel [1\bar{1}0]$ for (113)-oriented samples
(Fig.~\ref{fig11}) are in very good agreement with the
phenomenological expressions Eqs.~(\ref{equ14}), (\ref{equ16}),
and (\ref{equ18}) for $C_{2v}$ and Eqs.~(\ref{equ15}),
(\ref{equ17}) and (\ref{equ18}) for $C_s$ symmetry. Both figures
show experimental data  compared to calculations which were fitted
with one ordinate scaling parameter. The fact that $j_x$ is an
even function of $\Theta_0$ for (113)-oriented samples means that
in the sample under study the component $\gamma_{xz^\prime}$ in
Eqs.~(\ref{equ15})    of {\boldmath$\gamma$} is much larger
compared to $\gamma_{xy^\prime}$.

Microscopically CPGE can be the result of different optical
absorption mechanisms like inter-band transitions, inter-subband
transitions in QWs, Drude absorption etc. The CPGE at inter-band
absorption (valence band to conduction band) has  not been
observed experimentally so far. A strong spurious photocurrent due
to other mechanisms like the Dember effect, photovoltaic effects
at contacts  etc. mask the relatively weak CPGE. However
application of polarization selective measurements, like
modulation of polarization, should allow to extract the CPGE
current. In the infrared range, where effects mentioned above
vanish, the CPGE has been observed experimentally. In quantum well
structures absorption of infrared radiation may occur at indirect
intra-subband optical transitions (Drude absorption) and, for
photon energies being in resonance with the energy distance
between size quantized subbands, by direct transitions between
these subbands.

\subsubsection{Inter-subband  transitions in $n$-type QWs}
\label{IVA2}

\begin{figure}
\centerline{\epsfxsize 70mm \epsfbox{ 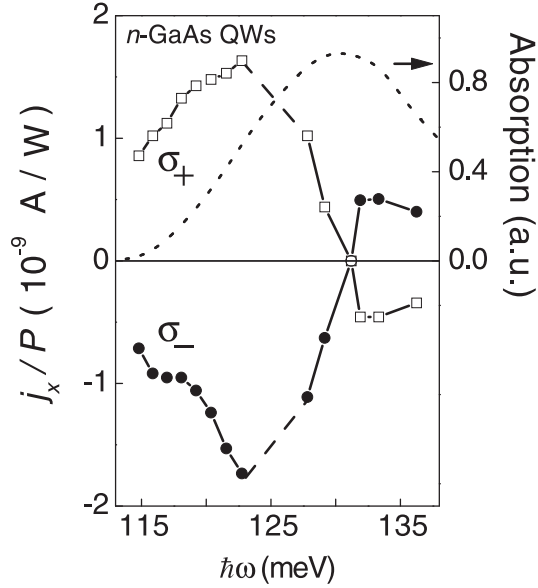 }}
\caption{Photocurrent in QWs normalized by $P$ as a function of
the photon energy $\hbar \omega$. Measurements are presented for
$n$-type (001)-grown GaAs/AlGaAs QWs of 8.2 nm width (symmetry
class C$_{2v}$) at $T=~$293~K and oblique incidence of  radiation
with an angle of incidence $\Theta_0 = 20^\circ$. The absorption
of the MIR laser radiation results in direct transitions between
$e${\it 1}  and $e${\it 2} subbands. The current $j_{x }$ is
perpendicular to the direction of light propagation. The dotted
line shows the absorption measured using a Fourier spectrometer. }
\label{fig12}
\end{figure}

Absorption of radiation in the range of 9~$\mu$m up to 11~$\mu$m
in $n$-type GaAs samples of QW widths  8.2~nm and 8.6~nm is
dominated by resonant direct inter-subband optical transitions
between the first and the second size-quantized subband.
Fig.~\ref{fig12} shows the resonance behaviour of the absorption
measured in GaAs QWs obtained by Fourier spectroscopy using  a
multiple-reflection waveguide geometry. Applying MIR radiation of
the CO$_2$ laser, which causes direct transitions in GaAs QWs, a
current signal proportional to the heli\-city $P_{circ}$ has been
observed at normal incidence in (113)-samples and at oblique
incidence in (001)-oriented samples indicating the spin
orientation induced circular photogalvanic effect~\cite{PRB03inv}.
In Fig.~\ref{fig12} the data are presented for a (001)-grown
$n$-GaAs QW of 8.2~nm width measured at room temperature. It is
seen that  the current for both, left and right handed circular
polarization, changes sign at a frequency $\omega = \omega_{inv}$.
This inversion frequency $\omega_{inv}$ coincides with the
frequency of the absorption peak. The peak frequency and
$\omega_{inv}$ depend on the sample width in agreement to the
variation of the subband energy separation. Experimental results
 shown in Fig.~\ref{fig12}, in particular the
sign inversion of the spectral behaviour of the current, are in a
good agreement with microscopic theory developed
in~\cite{PRB03inv} (see Eqs.~(\ref{equ33}) and (\ref{equ33a})).
The spectral sign inversion of the CPGE has also been detected in
a (113)-oriented $n$-GaAs QW which belongs to the point group
C$_s$. In this case the helicity dependent signal is observed in
$x$-direction at normal incidence of radiation along $z^\prime$.

%

The inversion of photon helicity driven current is a direct
consequence of {\boldmath $k$}-linear terms in the band structure
of subbands together with energy and momentum conservation as well
as optical selection rules for direct optical transitions between
size quantized subbands~\cite{PRB03inv}. At large photon energy,
$\hbar\omega > \varepsilon_{21}$, and  for QWs of $C_s$ symmetry
excitation occurs at positive $k_x$ resulting in a current $j_x$
shown by an arrow in Fig.~\ref{fig02}b. Decreasing of the photon
frequency shifts the transition towards negative $k_x$ and
reverses the direction of the current. In the frame of this model
the inversion of sign of the current takes place  at the photon
energy $\hbar \omega_{inv}$ corresponding to optical transitions
from the band minima. This shift of $\omega_{inv}$ away from the
frequency of peak absorption cannot be resolved in experiment on
currently available samples because of the broadening of the
absorption~\cite{PRB03inv}. Similar arguments hold for C$_{2v}$
symmetry (relevant for (001)-oriented samples) under oblique
incidence (see Fig.~\ref{fig05}a) although the simple selection
rules are no longer valid~\cite{book,Warburton96p7903}. Due to
selection rules the absorption of circularly polarized radiation
is spin-conserving but the rates of inter-subband transitions are
different for electrons with the spin oriented co-parallel and
anti-parallel to the in-plane direction of light
propagation~\cite{IT}. The asymmetric distribution of
photo-excited electrons results in a current which is caused by
these spin-conserving but spin-dependent
transitions~\cite{PRB03inv}.

\subsubsection{Inter-subband  transitions in $p$-type QWs}
\label{IVA3}

The  helicity dependent current of the spin orientation induced
CPGE has also been observed in $p$-type GaAs QWs due to
transitions between heavy-hole ($hh${\it 1}) and light-hole
($lh${\it 1}) subbands demonstrating  spin orientation of holes
(see Fig.~\ref{fig10}, lower plate)~\cite{APL00,PRL01,PRL02}. QWs
with various widths in the range of 4 to 20~nm were investigated.
For direct inter-subband transitions photon energies between
35~meV and 8~meV of FIR radiation corresponding to these QW widths
were applied. Due to  different effective masses of light and
heavy holes the absorption does not show narrow resonances.
Cooling the sample from room temperature to 4.2~K leads to a
change of  sign of spin orientation induced CPGE but the
$\sin2\varphi$ dependence is retained~\cite{APL00}. This
temperature dependent change of sign of the photogalvanic current,
which was also observed in $n$-type samples at direct transitions,
may be caused by the change of scattering mechanism from impurity
scattering  to phonon assisted scattering (see
section~\ref{IIB3}).

\subsubsection{Intra-subband  transitions in QWs}
\label{IVA4}

Optical absorption caused by indirect transitions in $n$-type
samples have been obtained applying FIR radiation covering the
range of 76~$\mu$m to 280~$\mu$m corresponding to photon energies
from 16~meV to 4.4~meV. The experiments were carried out on
GaAs~\cite{PRL01,PhysicaE02},  InAs~\cite{PRL01} and semimagnetic
ZnBeMnSe~\cite{DPG01} QWs. The energy separation between $e${\it
1} and $e${\it 2} size-quantized subbands of those samples is much
larger than the FIR photon energies used here. Therefore the
absorption is caused by indirect intra-subband optical
transitions. With illumination of (001)-grown QWs at oblique
incidence of FIR radiation a current signal proportional to the
helicity $P_{circ}$ has been observed (see Fig.~\ref{fig10}, upper
plate) showing that Drude absorption of a 2D electron gas results
in spin orientation and the CPGE. Spin orientation induced CPGE at
intra-subband absorption was also observed in $p$-type samples at
long wavelengths~\cite{APL00,PhysicaE02,PRB02}, where the photon
energies are smaller than the energy separation between the first
heavy-hole and the first light-hole subbands.

%
\begin{figure}
\centerline{\epsfxsize 86mm \epsfbox{ 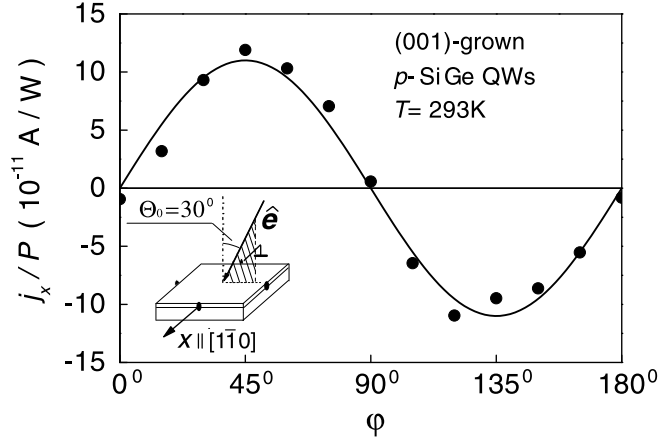 }}
\caption{Photogalvanic current $j_x$ normalized by $P$ in
(001)-grown and asymmetrically doped SiGe QWs and measured at room
temperature as a function of the phase angle $\varphi$. The data
were obtained under oblique incidence of irradiation at $\lambda =
10.6~\mu$m. The full line is fitted after Eqs.~\protect
(\ref{equ14}). The inset shows the geometry of the experiment.}
\label{fig13}
\end{figure}

%
\begin{figure}[h]
\centerline{\epsfxsize 76mm \epsfbox{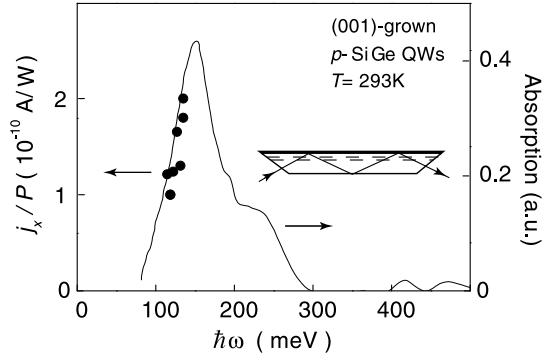  }}
\caption{Spectral dependence of spin orientation induced CPGE
(full dots) in (001)-grown and asymmetrically doped SiGe QWs due
to direct transitions between $hh${\it 1} and $lh${\it 1} valence
subbands at room temperature. The full line shows the absorption
spectrum obtained at 10~K. The absorption has been determined by
transmission measurements making use of a multiple-reflection
waveguide geometry shown in the inset. }
\label{fig14}
\end{figure}

\subsubsection{Spin orientation induced CPGE in SiGe  QWs}
\label{IVA5}

In symmetrically (001)-grown and symmetrically doped SiGe QWs no
photogalvanic current has been observed as expected from the
presence of inversion symmetry in both materials. However, in
non-symmetric QWs as described in section~\ref{IIIA}, spin
orientation induced CPGE has been observed being caused by the
Rashba spin-orbit coupling due to built-in potential gradient in
the QWs~\cite{PRB02,PASPS02sige}. Spin orientation induced CPGE is
most clearly seen at $hh${\it 1}-$lh${\it 1} inter-subband
absorption in (001)-oriented $p$-type QWs. With illumination by
MIR radiation of the CO$_2$ laser a current signal proportional to
the helicity $P_{circ}$ is observed under oblique
incidence~(Fig.~\ref{fig13}). For irradiation along $\langle 110
\rangle$ as well as along $\langle 100 \rangle$ crystallographic
directions the photocurrent flows perpendicular to the propagation
direction  of the incident light. Therefore only a transverse CPGE
was observed. It means that the effect  of a Dresselhaus-like
{\boldmath$k$}-linear term yielding a longitudinal effect for
$\langle 100 \rangle$ is negligible~\cite{PASPS02sige}. The
wavelength dependence of the photocurrent obtained between
9.2~$\mu$m and 10.6~$\mu$m corresponds to the spectral behaviour
of direct inter-subband absorption between the lowest heavy-hole
and light-hole subbands measured in transmission~(see
Fig.~\ref{fig14}).

\begin{figure}[h]
\centerline{\epsfxsize 86mm \epsfbox{ 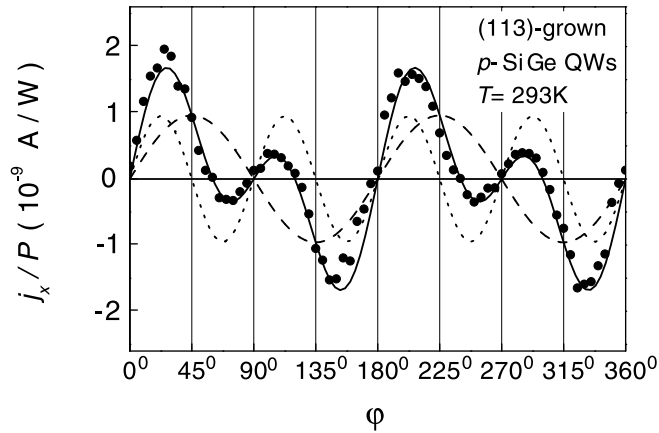 }}
\caption{Photogalvanic current in (113)-grown SiGe QWs normalized
by the light power $P$ as a function of the phase angle $\varphi$.
The results were obtained at  $\lambda = 280~\mu$m under normal
incidence of irradiation at room temperature. The full line is
fitted after left equations of Eqs.~\protect (\ref{equ15})
and~(42) corresponding to CPGE and LPGE, respectively. Broken and
dotted lines show $j_x \propto \sin 2\varphi$ and $j_x \propto
\sin 2\varphi \cdot \cos 2 \varphi$, respectively.}
\label{fig15}
\end{figure}

In the FIR range a more complicated dependence of the current as a
function of helicity has been observed. In (001)-grown asymmetric
quantum wells as well as in (113)-grown samples the dependence of
the current on the phase angle $\varphi$ may be described by the
sum of  two terms, one of them is $\propto\sin 2\varphi$ and the
other $\propto\sin2 \varphi \cdot \cos 2 \varphi$. In
Fig.~\ref{fig15} experimental data and a fit to these functions
are shown for a step bunched (001)-grown SiGe sample. The first
term is due to the spin orientation induced CPGE and the second
term is caused by a linear photogalvanic effect ~\cite{PRB02}
which will be discussed later (see section~5.1, Eqs.~(42). For
circularly polarized radiation the linear photogalvanic term $\sin
2 \varphi \cdot \cos 2\varphi$ is equal to zero and the observed
current is due to spin orientation induced CPGE only. We would
like to point out that, in agreement to symmetry,  the same term
may also be  present in zinc-blende structure based QWs but has
not yet been detected. CPGE and LPGE have different microscopic
physical mechanisms. Variation  of material parameters, excitation
wavelengths, and temperature may change the relative strengths of
these effects. For both spectral ranges, MIR and FIR, the angle of
incidence dependence of CPGE in SiGe structures is the same as
shown above for zinc-blende structure based materials.

%

The experimental results described so far are due to an imbalance
of photoexcited spin polarized electrons in {\boldmath$k$}-space
yielding the circular photogalvanic effect. After momentum
relaxation of the photoexcited carriers spin orientation induced
CPGE vanishes, however, a spin orientation may still be present if
the spin relaxation time is longer than the momentum relaxation
time. In such a case the spin-galvanic effect may contribute to
the total current.  In the next section we present experimental
results  demonstrating a pure spin-galvanic effect without
admixture of spin orientation induced CPGE.

\newpage

\begin{center}

{\bf {\Large  please replace the list of references  by part II\\
starting with the page number 28}}
\end{center}

\section*{Acknowledgement}

The authors thank  E.L.~Ivchenko, V.V.~Bel'kov,  L.E.~Golub,
S.N.~Danilov and Petra~Schneider for many discussions and helpful
comments on the present manuscript. We are also indebted to
G.~Abstreiter, V.V.~Bel'kov, M.~Bichler, J.~DeBoeck, G.~Borghs,
K.~Brunner, S.N.~Danilov, J.~Eroms, E.L.~Ivchenko, S.~Giglberger,
P. Grabs, L.E.~Golub, T.~Humbs, J.~Kainz, H.~Ketterl, B.N.~Murdin,
Petra~Schneider, D.~Schuh, M.~Sollinger, S.A.~Tarasenko,
L.~Molenkamp, R.~Newmann, V.I.~Perel, C.R.~Pidgeon,
 P.J.~Phillips, U.~R\"{o}ssler, W.~Schoepe, D.~Schowalter, G. Schmidt, V.M.~Ustinov, L.E.~Vorobjev,
 D.~Weiss, W.~Wegscheider, D.R.~Yakovlev, I.N.~Yassievich, and A.E.~Zhukov,  for long
standing cooperation during the work on spin photocurrents. We
gratefully acknowledge financial support by the Deutsche
Forschungsgemeinschaft (DFG), the Russian Foundation for
Basic Research (RFBR) and the NATO Linkage Grant which
made this work possible.




\end{document}